\documentclass[aps,onecolumn,preprintnumbers,pra]{revtex4-2}
\usepackage{caption}
\usepackage{tabularx}
\usepackage{graphicx}  
\usepackage{subfigure}
\usepackage{multirow}
\usepackage{mathtools}
\usepackage{notes2bib}
\usepackage{tikz}
\usepackage{float}

 \Large

\usepackage[utf8]{inputenc}
\usepackage[english]{babel}
\usepackage{tikz}
\usetikzlibrary{shapes.geometric}

\usepackage{fancyhdr}
\usepackage{parskip}
\usepackage[T1]{fontenc}
\usepackage{dcolumn}   

\usepackage{bm}        
\usepackage{amsfonts}  
\usepackage{amsmath}   
\usepackage{amssymb}   

\newcommand{\pwisein}{\left\{ \begin{array}{ll}}
\newcommand{\pwiseout}{\end{array}\right.}

\usepackage[unicode=true,pdfusetitle,
 bookmarks=true,bookmarksnumbered=false,bookmarksopen=false,
 breaklinks=false,pdfborder={0 0 0},pdfborderstyle={},backref=false,colorlinks=true]
 {hyperref}
\hypersetup{allcolors=blue}

\makeatletter
\renewcommand\NAT@citesuper[3]{\ifNAT@swa
\unskip\hspace{1\p@}\textsuperscript{[#1]}%
\if\relax#3\relax\else\ [#3]\fi\else [#1]\fi\endgroup}
\makeatother

\usepackage [english]{babel}
\usepackage [autostyle, english = american]{csquotes}
\MakeOuterQuote{"}

\begin{document}
\setcitestyle{super}
\title{Quantum Pointwise Dimension (QUPID): An algorithm for identifying classical chaos in quantum dynamics}

\author{Jonathan Berkheim, David J. Tannor}
\affiliation {\it Department of Chemical and Biological Physics, Weizmann Institute of Science, 76100, Rehovot, Israel}

\begin{abstract}  
The Pointwise Dimension is a diagnostic that reveals the effective dimensionality of classical trajectories in a $2N$-dimensional phase space. The essence of this diagnostic is to find a scaling according to which density changes with radius $r$ around a point of interest in phase space. Particularly, regular trajectories are confined to an $N$-dimensional invariant torus, while chaotic trajectories explore a $(2N-1)$-dimensional energy surface. In classical mechanics, this trajectory density can be found by counting recurrences to within the sphere of radius $r$. In quantum mechanics, probability density is intrinsic, making the Pointwise Dimension compelling, but surprisingly this has not been studied in the quantum regime. In this work, we implement the Pointwise Dimension in the quantum regime using the Husimi representation of the quantum wavepacket. This implementation entails a long-time average of the time-dependent Husimi representation to obtain a stationary phase space density, and a radial spatial integration of this density over a sphere of radius $r$. The classical-quantum correspondence is excellent for both autonomous and non-autonomous systems, adding a new diagnostic to the study of quantum chaos. Finally, we explore the possibility, both classically and quantum mechanically, of obtaining the Pointwise Dimension from a time correlation function.
\end{abstract}

\maketitle 
\section{Introduction}
\begin{figure}[ht!]
    \centering
    \includegraphics[width=0.8\linewidth]{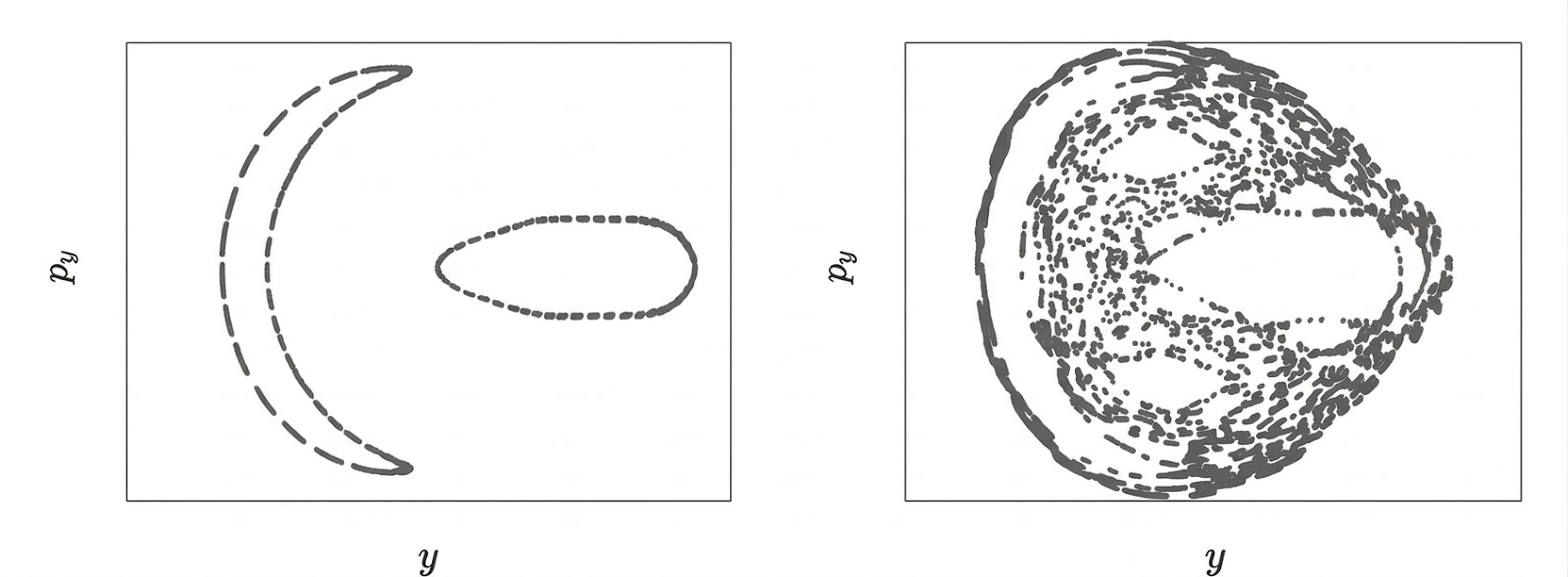}
    \caption{Poincaré surfaces of section for two representative trajectories in the Hénon-Heiles potential, $V(x,y)=\frac{1}{2}(x^2+y^2)+\lambda(x^2y-\frac{1}{3}y^3)$, where $\lambda=1/\sqrt{80}$ and $E=1/8$: regular trajectory (left) and chaotic trajectory (right) emerge from two different sets of initial conditions in phase space. The geometrical difference between two behaviors is seen as the regular trajectory assumes a well-ordered locus, whereas the chaotic trajectory spreads over a scattered sea.}
    \label{fig:Poincare}
\end{figure}

Classical trajectories live in a $2N$-dimensional phase space, where $N$ is the number of spatial degrees of freedom. In certain cases, called regular or integrable, there are $N$ conserved quantities so that the trajectory lives on an $N$-dimensional invariant torus in the $2N$-dimensional phase space. For chaotic motion, equations of motion (EOMs) are non-integrable; the torus structure hardly exists or does not exist at all. The only constant of motion in the autonomous case is the total energy; thus, the trajectory lives in a $(2N-1)$-dimensional surface in phase space. This profound geometrical difference between regular and chaotic motion, which Einstein noted,\citep{Einstein1917OnEpstein} lies at the core of many tools for distinguishing between the dynamical behaviors of trajectories. E.g., it is seen in the familiar Poincaré surfaces of section, where regular trajectories appear as closed loci, whereas chaotic trajectories appear as a sea of scattered points.\citep{Gutzwiller1990ChaosMechanics,Lichtenberg1992RegularDynamics} Figure \ref{fig:Poincare} illustrates this difference using two representative trajectories in the Hénon-Heiles potential.

The Pointwise Dimension is a fractal method aimed at evaluating the effective dimensionality of the trajectories and, therefore, can be used to distinguish between regular and chaotic classical trajectories. This dimension is based on the local scaling of the density enclosed within spheres of decreasing radii; it corresponds to the power law that characterizes the polynomial scaling of the density as the radii approach zero. That said, the notion of density is somewhat alien to classical mechanics when a single trajectory is considered; thus, it must be implemented artificially, as discussed below. On the contrary, in quantum mechanics, probability density is very fundamental. And, as similar questions regarding regularity and ergodicity can be asked about the time evolution of a quantum wavefunction, we would like to use the probability density to resolve the effective dimensionality of the quantum dynamics. 
There has been a long-standing dispute on whether quantum chaos exists, in the sense that exponential sensitivity to initial state does not exist in quantum mechanics. It was Michael Berry who drew the distinction, suggesting the term "quantum chaology" to describe the rich phenomena linking classical chaos to quantum mechanics.\citep{Berry1989QuantumChaos}
But, while ergodicity of quantum systems has been extensively studied for decades, e.g., using random matrix theory, there is much less literature about the quantum chaos of moving wavefunctions. 

In this work, we will formulate the Pointwise Dimension method in the quantum framework and apply it to several systems. Particularly, in the semiclassical limit $\hbar\to0$, the results show remarkable agreement with the classical ones. Our case studies are (1) $1D$ autonomous systems: simple harmonic oscillator and quartic oscillator; (2) $1D$ non-autonomous, time-periodic system, the Kicked Rotor; (3) $2D$ autonomous systems: simple harmonic oscillator and quartic coupled oscillator.

\section{Background: Pointwise Dimension for classical trajectories}
The definition of the local Pointwise Dimension $d_{c}$ for classical systems, evaluated at a reference phase space point $\textbf{z}_{0}=(\textbf{x}_{0},\textbf{p}_{0})$ is\citep{Billingsley1999ConvergenceMeasures}
\begin{equation}
\label{pointwise-def}
d_{c}(\textbf{z}_{0})=\lim_{r\to0}\frac{\log \mu(S_{r}(\textbf{z}_{0}))}{\log r},
\end{equation}
where $S_{r}$ is a sphere of radius $r$, centered at $\textbf{z}_{0}$; $\mu$ is a density measure quantifying the "probability" of finding the trajectory inside the sphere $S_{r}$. A natural choice of $\textbf{z}_{0}$ would be the starting point of the trajectory, $\boldsymbol{\zeta}_{0}$, but any point along the trajectory is valid as well. This definition is consistent with Mandelbrot's original formulation of Fractal Dimension,\citep{Mandelbrot1967HowDimension} extending the logic of scaling and self-similarity to the local density of a trajectory in phase space. However, the formal definition in eq. \ref{pointwise-def} does not have much numerical meaning, since $\mu$, the density measure, is not well-defined for a single classical trajectory. To overcome this, Martens and Ezra\citep{Martens1984ADimension} proposed an easy-to-implement algorithm that uses time-series as the primary entity. In their algorithm, the first step is to obtain the time-dependent Euclidean norm of the trajectory $\textbf{z}(t)=(\textbf{x}(t),\textbf{p}(t))$, defined with respect to the reference point $\textbf{z}_{0}$:
\begin{equation}
\label{dist-time-series}
D(t) =||\textbf{z}(t)-\textbf{z}_{0}||=\sqrt{\sum_{{i}=1} ^{N}(x_{i}(t)-x_{i,0})^2+\left(\frac{\omega_{i}}{m}\right)^2(p_{i}(t)-p_{i,0})^2},
\end{equation}
where $i$ runs over degrees of freedom. The constant weights $\omega_{i}/m$ scale the units of momentum to match those of position, and in general, they can be taken as unity, as we do in this work. The examples presented by Martens and Ezra were essentially invariant to the values of $\omega_{i}/m$.\citep{Martens1984ADimension} Then, sort $D(t)$ from smallest to largest values, which rearranges the distances by spatial proximity to $\textbf{z}_{0}$. Returning to eq. \ref{pointwise-def}, we define the density measure $\mu(S_{r}(\textbf{z}_{0}))$ as $\mathcal{N}_{c}(r,\textbf{z}_{0})$, the accumulated density up to radius $r$ around $\textbf{z}_{0}$:
\begin{equation}
\label{theta}
\mu(S_{r}(\textbf{z}_{0}))=\mathcal{N}_{c}(r,\textbf{z}_{0})=\int_{0}^{r}\sum_{i}\delta(D(t_{i})-r)dr=\sum_{i}\Theta(r-D(t_{i})),
\end{equation}
where $t_{i}$ are the recurrence times, i.e., moments where the sorted function $D$ intersects with a certain shell $r$, and $\delta$ and $\Theta$ are the Dirac delta function and the Heaviside step function, respectively. This quantity, $\mathcal{N}_{c}(r,\textbf{z}_{0})$, estimates the average number of recurrences within a sphere at radius $r$ around $\textbf{z}_{0}$. The Pointwise Dimension corresponds to the slope of the linear part (i.e., at small values of $r$) of the curve of $\log \mathcal{N}_{c}(r,\textbf{z}_{0})$ vs. $\log r$. Martens and Ezra claimed\citep{Martens1984ADimension} that the dependence on $\textbf{z}_{0}$ might be removed, i.e., that classical trajectories are essentially monofractals. We find that this statement requires caution, unless $d_{c}$ is rounded to the closest integer; practically, statistics is improved when several $\textbf{z}_{0}$ are considered and the density is averaged over several reference points $\textbf{z}_{0}=\{\textbf{z}_{0}^{(1)},\textbf{z}_{0}^{(2)},...,\textbf{z}_{0}^{(M)}\}$. One rather defines
\begin{equation}
\label{Nrz-cl-av}
\mathcal{N}_{c}(r)=\frac{1}{M}{\sum_{i=1}^{M} \mathcal{N}_{c}\left(r,\textbf{z}_{0}^{(i)}\right)};\hspace{0.5cm}d_{c}\left(\boldsymbol{\zeta}_{0}\right)=\lim_{r\to0}\frac{\log \mathcal{N}_{c}(r)}{\log r}=\lim_{r\to0}\frac{\log \langle\mu(S_{r}(\boldsymbol{\zeta}_{0}))\rangle}{\log r},
\end{equation}
where $\langle\mu(S_{r}(\boldsymbol{\zeta}_{0}))\rangle$ denotes the averaged density measure over the reference points $\textbf{z}_{0}$ associated with the initial condition $\boldsymbol{\zeta}_{0}$. The dependence on $\boldsymbol{\zeta}_{0}$ highlights the crucial role of the starting point in dictating regular versus chaotic motion. 

As mentioned, the main goal of this work is to implement the Pointwise Dimension method in quantum systems to investigate the signature of classical chaos in the corresponding quantum dynamics. To this end, we also need to define $\mu$ in the quantum world.
\begin{figure}
    \centering
    \includegraphics[width=0.5\linewidth]{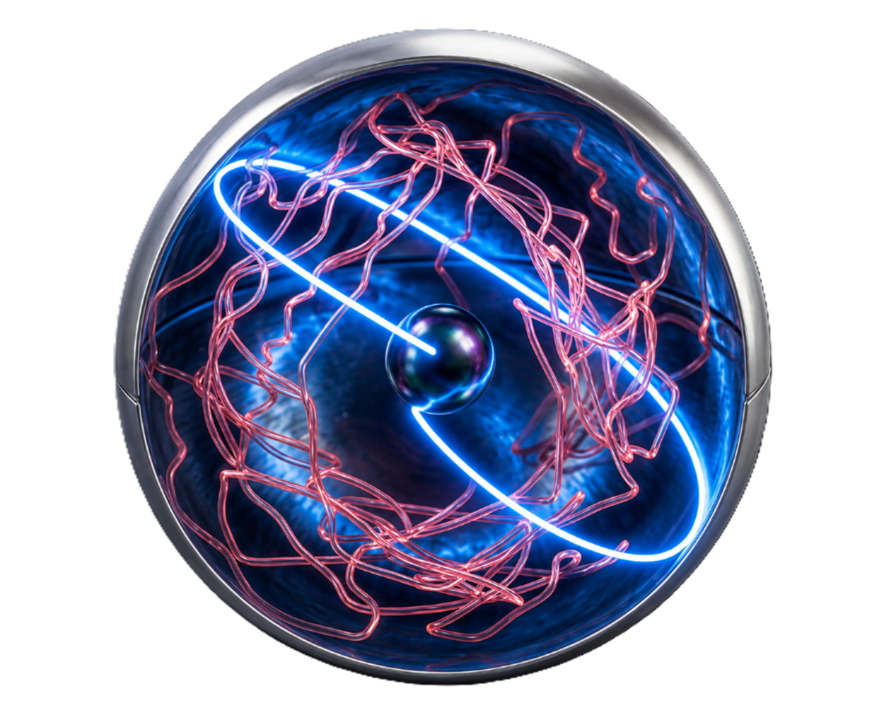}
    \caption{Pictorial representation of the Pointwise Dimension. The regular trajectory (blue) is confined to a compact structure, while the chaotic trajectory (red) is spread over the entire volume. The difference between behaviors is seen in the recurrence of the regular trajectory and the absence of the chaotic trajectory within the small sphere, respectively.}
    \label{fig:pointwise_artwork}
\end{figure}

\section{Quantum Analog of the Pointwise Dimension}
\subsection{Theory}
There are several definitions for the distance between quantum states, such as the Hilbert-Schmidt distance and the Bures distance,\citep{Bengtsson2006GeometryStates} but none of them is expected to correspond to classical phase space distance, even in the semiclassical limit. For example, the Hilbert-Schmidt distance is based on the time-autocorrelation function, i.e., for pure states, $C(t)=\langle\Psi(0)|\Psi(t)\rangle$: 
\begin{equation}
\label{AC}
D(t)=\sqrt{2(1-{|C(t)|^2} )},
\end{equation}
and the Bures distance is the same, but without the power of $2$. These distances, as well as others, are commonly used in quantum information theory, where distinguishability of states is discussed. However, such overlap-based distances do not provide a consistent geometric phase-space interpretation, since $C(t)$ is calculated in Hilbert space where the overall phase differences are interpreted as distances. This statement is correct even in the semiclassical limit, with the exception of the simple harmonic oscillator, due to the exact preservation of the gaussian form of coherent states without dispersion.

With the understanding that autocorrelation-based distance is inadequate for our needs, we go the other way around. Instead of seeking quantal recurrences and a time-series, we return to the definition in eq. \ref{pointwise-def} and attempt to employ a phase-space density measure directly. Regular vs. chaotic dynamics is usually evaluated at long timescales, so that we are interested in an asymptotic density
\begin{equation}
\mathcal{N}_{q}(\textbf{z})=\lim_{\tilde{T}\to\infty}\frac{1}{\tilde{T}}\int_{0}^{\tilde{T}}f(\textbf{z},t)dt,\end{equation}
where $f(\textbf{z},t)$ is the candidate time-dependent density distribution and $\tilde{T}$ is the termination time of the dynamics. 

The immediate candidate for such a density is the Wigner representation, $W(\textbf{z},t)$,\citep{Wigner1932OnEquilibrium} a $2N$-dimensional phase-space representation of the quantum wavefunction $\Psi(\textbf{x},t)$:
\begin{equation}
\label{Wigner-def}
W(\textbf{z},t)=\frac{1}{(2\pi)^{N}}\int_{-\infty}^{\infty}\Psi^{*}\left(\textbf{x}-\frac{\textbf{s}}{2},t\right)\Psi\left(\textbf{x}+\frac {\textbf{s}}{2},t\right)e^{-i\textbf{p}\cdot\textbf{s}/\hbar}d\textbf{s},
\end{equation}
where $\textbf{s}$ is an auxiliary coordinate, conjugate to $\textbf{p}$. The probabilities in the position and momentum spaces can be read off as marginal distributions of the Wigner representation, which provides a qualitative picture for the concentration of probability in phase space. However, this representation suffers from a major drawback: it is not positive-definite. As a result, the Wigner representation is not a good candidate, as the accumulation over time might result in regions of destructive interference between $f(\textbf{z},t_{1})=W(\textbf{z},t_{1})>0$ and $f(\textbf{z},t_{2})=W(\textbf{z},t_{2})<0$. Such regions can compromise the clean polynomial scaling of the density vs. radius.

As an alternative to the Wigner representation, we consider the Husimi representation,\citep{Husimi1940SomeMatrix} defined as an expectation value of the density operator $\hat\rho(t)$ in the coherent state basis:
\begin{equation}
\label{Husimi-def}
Q(\textbf{z}',t)=\frac{1}{\pi}\langle\alpha_{\textbf{z}'}|\hat\rho(t)|\alpha_{\textbf{z}'}\rangle,
\end{equation}
where $|\alpha_{\textbf{z}'}\rangle$ is the coherent state basis centered at $\textbf{z}'=(\textbf{x}',\textbf{p}')$, 
\begin{equation}
\label{gaussian}
\langle\textbf{x}|\alpha_{\textbf{z}'}\rangle=g_{\alpha}(\textbf{x};\textbf{z}')=\left(\frac{2\alpha}{\pi}\right)^{N/4}\exp\left[-\alpha(\textbf{x}-\textbf{x}')^{2}+\frac{i}{\hbar}\textbf{p}'(\textbf{x}-\textbf{x}')+\frac{i}{\hbar}\gamma\right],
\end{equation}
$\alpha=m\omega/2\hbar$ and $\gamma$ is a global phase factor. The Husimi representation can also be seen as a gaussian smoothing of the Wigner representation. In contrast to the Wigner representation, the Husimi representation is positive-definite and thus can serve as a faithful phase-space density. For pure states, we can write
\begin{equation}
\label{Husimi-abbv}
Q(\textbf{z}',t)=\frac{1}{\pi}|\langle\alpha_{\textbf{z}'}|\Psi(t)\rangle|^{2}.
\end{equation}
This representation has been successfully employed in several works on the quantum chaos of eigenstates.\citep{Weissman1982QuantumStates,Arranz2013OnsetFunction,Wustmann2010StatisticalSystems,Heller2018TheSpectroscopy} For simplicity, we will drop the $1/\pi $ normalization henceforth.

Based on classical-quantum correspondence, we claim that in the limits $\hbar\to0$ and $\tilde T\to\infty$ the Husimi representation should encode information about the classical trajectory starting from the phase space point $\boldsymbol{\zeta}_{0}$ where $\Psi(\textbf{x},0)$ is centered. In this work, we consider $\Psi(\textbf{x},0)$ as a coherent state as well (eq. \ref{gaussian}). To formulate the Quantum Pointwise Dimension, our suggested methodology relies on two main steps: (1) evaluating the long-time average of the Husimi representation, playing the role of the quantum phase space density, and (2) integrating the stationary density over a sphere of radius $r$ centered at a reference point $\textbf{z}_{0}$.

Inserting the completeness relation $I=\sum_{n}|\phi_{n}\rangle\langle\phi_{n}|$ in eq. \ref{Husimi-abbv}, modifying a derivation by Heller,\citep{Heller2018TheSpectroscopy} one can show that for non-degenerate quantum systems
\begin{multline}
\label{Husimi-inf}
Q_{\infty}(\textbf{z}')=\lim_{\tilde{T}\to\infty}\frac{1}{\tilde{T}}\int_{0}^{\tilde{T}}Q(\textbf{z}',t)dt=\lim_{\tilde{T}\to\infty}\frac{1}{\tilde{T}}\int_{0}^{\tilde{T}}|\langle\alpha_{\textbf{z}'}|\Psi(t)\rangle|^{2} dt=\lim_{\tilde{T}\to\infty}\frac{1}{\tilde{T}}\int_{0}^{\tilde{T}}\sum_{n,m}\langle\alpha_{\textbf{z}'}|\phi_{n}\rangle\langle\phi_{n}|\Psi(t)\rangle\langle \Psi(t)|\phi_{m}\rangle\langle\phi_{m}|\alpha_{\textbf{z}'}\rangle dt=\\
\lim_{\tilde{T}\to\infty}{\left[\underbrace{\sum_{n\neq m} \langle\alpha_{\textbf{z}'}|\phi_{n}\rangle\langle\phi_{m}|\alpha_{\textbf{z}'}\rangle c_{m}^{*}c_{n}e^{-i(E_{n}-E_{m})\tilde{T}/2\hbar}\text{sinc}\left(\frac{E_{n}-E_{m}}{2\hbar}\tilde{T}\right)}_{\text{non-resonant terms}}+\underbrace{\sum_{n}|c_{n}|^{2}|\langle \alpha_{\textbf{z}'}|\phi_{n}\rangle|^{2}}_{\text{resonant terms}}\right]}=\sum_{n=1}^{\infty}|c_{n}|^{2}Q_{n}(\textbf{z}'),
\end{multline}
where $c_{n}=\langle\phi_{n}|\Psi(0)\rangle$ is $n$-th spectral coefficient of $\Psi(0)$ and $Q_{n}(\textbf{z}')$ is the Husimi representation of eigenstate $|\phi_{n}\rangle$at $\textbf{z}'$. Specifically, for a non-degenerate system, where $E_{n}\neq E_{m}$ for all $n,m$, the argument of the $\text{sinc}$ function is nonzero; all the non-resonant terms undergo phase cancellation over long-time averaging and the state $\Psi(0)$ is projected onto the time-independent diagonal ensemble. This result, $Q_{\infty}(\textbf{z}')=\sum_{n}|c_{n}|^{2}Q_{n}(\textbf{z}')$, has also been used in the study of eigenstate thermalization \citep{Deutsch1991QuantumSystem,Srednicki1994ChaosThermalization,Rigol2012AlternativesThermalization} applied to the phase space projection operator $\hat{O}=|\alpha_{\textbf{z}'}\rangle\langle\alpha_{\textbf{z}'}|$.

Equation \ref{Husimi-inf} can be written in the form
\begin{equation}
\label{Pab_spec}
Q_{\infty}(\textbf{z}')=P(\Psi(0)|\alpha_{\textbf{z}'})=\sum_{n}|\langle\phi_{n}|\Psi(0)\rangle|^{2}|\langle \alpha_{\textbf{z}'}|\phi_{n}\rangle|^{2}.
\end{equation}
where $P(\Psi(0)|\alpha_{\textbf{z}'})$ is recognized as a special case of $P(a|b)$.\citep{Heller1987QuantumSpace}

With this $Q_{\infty}(\textbf{z}')$ in hand we can directly obtain
\begin{equation}
\label{Nrz-q}
\mathcal{N}_{q}(r,\textbf{z}_{0})=\int_{\Omega} Q_{\infty}(\textbf{z}')d\textbf{z}'
\end{equation}
where $\Omega=\{\textbf{z}'|r_{\min}\leq||\textbf{z}'-\textbf{z}_{0}||\leq r\}$. The meaning of $r_{\min}$ is detailed below and $\textbf{z}_{0}=\{\textbf{z}_{0}^{(1)},\textbf{z}_{0}^{(2)},...,\textbf{z}_{0}^{(M)}\}$ is a random set of reference points drawn from the distribution $Q_{\infty}$. As in the classical case, one rather compute the reference-averaged version of $\mathcal{N}(r,\textbf{z}_{0})$, which is desirable in order to improve statistics:
\begin{equation}
\label{Nrz-q-av}
\mathcal{N}_{q}(r)=\frac{1}{M}{\sum_{i=1}^{M} \mathcal{N}_{q}\left(r,\textbf{z}_{0}^{(i)}\right)}.
\end{equation}
We conjecture that $Q(\textbf{z}')$ and consequently $\mathcal{N}_{q}(r)$ should reflect the dynamical behavior of the moving wavefunction. For regular dynamics, $Q_{\infty}(\textbf{z}')$ should localize on a compact structure resembling the classical invariant torus; for chaotic dynamics, $Q_{\infty}(\textbf{z}')$ should fill the available phase space uniformly. By analogy with the classical definition, we will define the Quantum Pointwise Dimension (QUPID):
\begin{equation}
\label{dq-def}
d_{q}(\boldsymbol{\zeta}_{0})=\lim_{r\to0}\frac{\log \mathcal{N}_{q}(r)}{\log r}=\lim_{r\to0}\frac{\log\langle\mu(S_{r}(\boldsymbol\zeta_{0}))\rangle}{\log r}.
\end{equation}
Thus, $d_{q}$ can be viewed as a generalization of Heller's $P(a|b)$ to measure the scaling with dimensionality as an indication of chaos.

This framework applies to autonomous systems, but can also be extended to non-autonomous, time-periodic systems via Floquet theory. We assume that eq. \ref{Husimi-inf} is still valid if $H(t)=H(t+T)$, even though energy is not conserved, but there could be effective invariants.\citep{Wustmann2010StatisticalSystems} To handle such systems, we replace $|\phi_{n}\rangle$ by the Floquet states $|F_{\nu}\rangle$, which satisfy the following eigenvalue problem
\begin{equation}
\label{Floquet}
\hat{U}(T,0)|F_{\nu}\rangle= e^{-i\hat{H}T/\hbar}|F_{\nu}\rangle=e^{-i\epsilon_{\nu}T/\hbar}|F_{\nu}\rangle,
\end{equation}
where $\epsilon_{\nu}$ are the quasi-energies.\citep{Santoro2019IntroductionNotes} This eigenvalue problem reflects the stroboscopic mapping $|\Psi(nT)\rangle =\hat{U}(nT,0)|\Psi(0)\rangle=\left[\hat{U}(T,0)\right]^{n}|\Psi(0)\rangle$ where $n\in\mathbb{N}$. The validity of eq. \ref{Husimi-inf} follows from the argument that long-time stroboscopic averaging eliminates non-resonant terms $e^{-i(\epsilon_{\nu}-\epsilon_{\nu'})T/\hbar}$ for $\epsilon_{\nu}\neq\epsilon_{\nu'}$. As long as a time-periodic continuous Hamiltonian problem, e.g., Kicked Rotor $V(x)=K\cos x$, is properly mapped to a discrete mapping, then the resulting classical and quantum dynamics should correspond to each other in the semiclassical limit. In this sense, we believe that time-periodic problems can be well-treated using the Pointwise Dimension method, as detailed in Sec. IV. 

\subsection{Numerical considerations}
\textit{Choice of $\hbar$ and grid parameters.} 
All semiclassical methods are based on the assertion that $\hbar\to0$ leads to classical-quantum correspondence. Our algorithm is also based on this assertion, since we are concerned with the classical ergodicity underlying the quantum dynamics; in particular, the fine details of $Q_{\infty}(\textbf{z}')$ are of central importance to us. They cannot be structured unless $\hbar$ is small, since in the deep quantum regime dispersion suppresses ergodicity. However, we do not scan $\hbar$ continuously; upon numerical convergence (see Appendix), we consider $\hbar=\pi/100$ for the $1D$ simulations and $\hbar=\pi/50$ for the $2D$ simulations. When grid methods are discussed, the selection of the value of $\hbar$ directly affects all other parameters required to obtain a converged simulation. According to Heisenberg's uncertainty principle, $\Delta x\Delta p=h=2\pi\hbar$, where $\hbar$ play the role of the minimal phase-space cell. Consequently, $\Delta x=2\pi\hbar/p_{\max}$, and let $A$ be the "box volume" of the simulation in phase space,
\begin{equation}
\label{grid}
A=L_{x}\cdot L_{p}=(N_{\text{states}}\Delta x)\cdot \left(\frac{2\pi\hbar}{\Delta x}\right)=2\pi\hbar N_{\text{states}},
\end{equation}
and $N_{\text{states}}\sim1/\hbar$ corresponds to the highest-energy eigenstate obtained from the diagonalization, eq. \ref{TISE}. Using $\hbar$, $\Delta x$ and $N_{\text{states}}$ we make the diagonalization in practice. In the next step, it depends on which approach we take: if $Q_{\infty}$ is fully constructed over the coherent state basis, then $\Delta x'$ and $\Delta p'$ should both be smaller than $\sqrt{\hbar}$, by the same Heisenberg's logic. In MC simulations, $Q_{\infty}$ is constructed only around certain reference points $\textbf{z}'$, then $\Delta x'$ and $\Delta p'$ do not exist, but convergence has to be verified with respect to the number of MC samples.

\textit{Hilbert space diagonalization and truncation.} The computational advantage inherent in eq. \ref{Husimi-inf} deserves emphasis. A direct time-propagation of $\Psi(0)$ up to $\tilde{T}$ requires a total number of time steps $N_{t}=\tilde{T}/\Delta t\sim\tilde{T} E_{\max}/\hbar$, which becomes very large in the semiclassical limit $\hbar\to 0$; by contrast, a single diagonalization of $\hat{H}$ with standard grid-based methods has a one-time cost independent of $\tilde{T}$. In this way, our approach yields a qualitative advantage over direct propagation and, more fundamentally, the limit $\tilde{T}\to\infty$ is treated more rigorously.

In eq. \ref{Husimi-inf}, note that no actual propagation of $\Psi(0)$ is required in order to obtain $Q_{\infty}(\textbf{z}')$; one simply solves the time-independent Schrödinger equation,
\begin{equation}
\label{TISE}
\hat{H}|\phi_{n}\rangle=E_{n}|\phi_{n}\rangle,
\end{equation}
which for low dimensional systems can be solved using a plethora of numerical methods, e.g., Finite-Difference Method (FDM) or Pseudo-Spectral Methods.\citep{Tannor2007IntroductionPerspective} In this work, we have employed the 5th-order FDM, and, in practice, Hilbert space in eq. \ref{Husimi-inf} is truncated at a high-energy eigenstate.

\begin{figure}[ht!]
    \centering
    \includegraphics[width=0.6\linewidth]{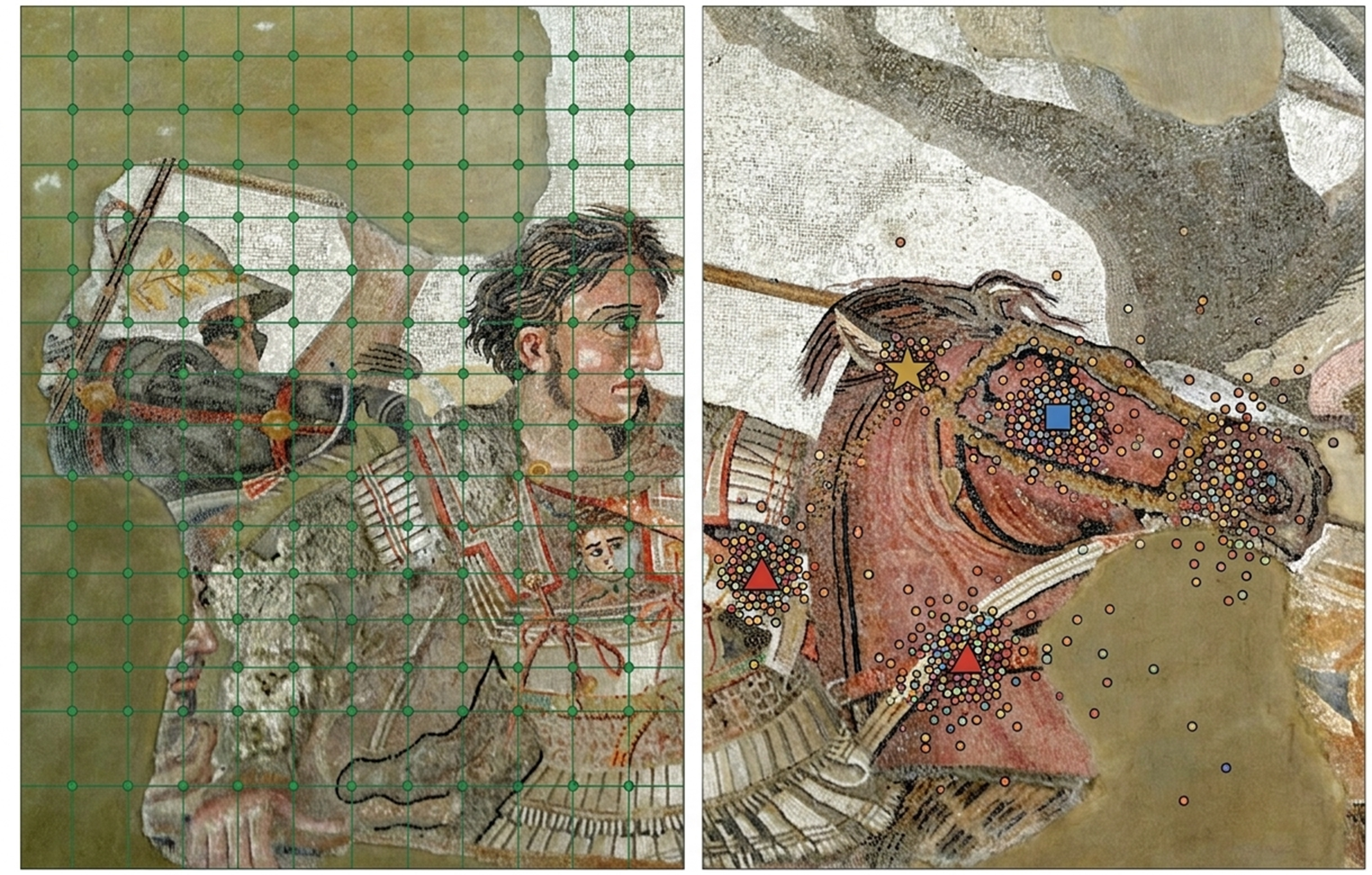}
    \caption{The main difference between grid sampling (left) and MC sampling (right). While the grid is systematic and uniform, it overlooks fine details. MC requires fewer reference points, but can resolve fine structures.}
    \label{fig:alexander}
\end{figure}

\textit{Grid vs. Monte-Carlo integration.} At this stage we notice that some multidimensional integrals in the procedure might cause a computational bottleneck; in particular, the integrals $c_{n}=\langle\phi_{n}|\Psi(0)\rangle$ and $Q_{n}=\langle\phi_{n}|\alpha_{\textbf{z}'}\rangle$ become too heavy in multidimensional systems. To avoid this notorious "curse of dimensionality", it is recommended to apply Monte-Carlo (MC) integration for at least part of these calculations. In contrast with grid-based methods which scale exponentially with the system's dimensionality, MC integration is based on random samplings where the error converges as $\mathcal{O}(1/\sqrt{N_{\text{samples}}})$ independently on the dimension, making it advantageous for multidimensional simulations. In general, MC methods are known to efficiently resolve fine structures and localized features, as demonstrated in fig. \ref{fig:alexander}.

\textit{Ancestral Sampling and state selection.} In an MC calculation of $c_{n}$, random samples are drawn from the gaussian distribution $|\Psi(0)|^{2}$, and, for the calculation of $Q_{n}$, the random samples are drawn from the phase-space distribution defined by the coherent state basis $|\alpha_{\textbf{z}'}\rangle$. In this work we use MC calculation of $Q_{n}$ using Ancestral Sampling.\citep{Bishop2007PatternLearning} In this framework, an eigenstate $|\phi_{n}\rangle$ is first selected according to its relative weight $|c_{n}|^{2}$, then a random point $\textbf{z}'$ is drawn from regions of nonzero density of that specific eigenstate. However, we calculate the $c_{n}$'s using a grid method. 

\textit{Discrete representation of Floquet operators.} For non-autonomous, time-periodic systems, the diagonalization of the evolution operator $U(T,0)$ is performed by constructing its discrete matrix representation, $\mathbf{U}$, via the Fourier pseudo-spectral method:\citep{Tannor2007IntroductionPerspective}
\begin{equation}
\label{pseudo-spectral}
\mathbf{U} = \text{diag}\left(e^{-\frac{i}{\hbar}V(x)T}\right) \cdot {\textbf{F}}^{-1} \cdot \text{diag}\left(e^{-\frac{i}{\hbar}\frac{p^{2}}{2m}T}\right) \cdot{\textbf{F}},
\end{equation}
where $\textbf{F}$ represents the Discrete Fourier Transform (DFT) matrix of size $N _{x}\times N_{x}$, and $N_{x}$ defines the sampling grid of the position space.

\section{Results and analysis}
In contrast to the classical Pointwise Dimension (eq. \ref{theta}), where the lower bound of the integral $\mathcal{N}_{q}(r)$ is $r=0$, in QUPID, the lower bound is $r_{\min}$ (eq. \ref{Nrz-q}). This $r_{\min}$, a small nonzero radius, is also associated with the value of $\hbar$; one must not consider $\hbar=0$, where quantum dynamics becomes ill-defined.\citep{Berry1989QuantumChaos} In the same sense, there exists an $r_{\max}$, a radius where $\mathcal{N}_{q}(r)$ reaches global scale truncation and the log-log plot plateaus. This truncation appears for both the classical and the quantum regimes.

\begin{figure}[ht!]
    \centering
    \includegraphics[width=0.3\linewidth]{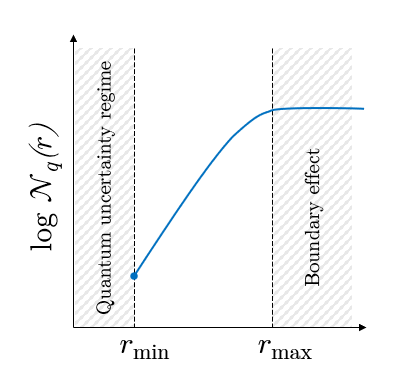}
    \caption{The conceptual meaning of $r_{\min}$ and $r_{\max}$ as the bounds of the quantum uncertainty regime and saturation regime, respectively. The slope is fitted within the unshaded area.}
    \label{fig:rmin_rmax}
\end{figure}

Altogether, we are limited to $r_{\max}\geq r\geq r_{\min}>\sqrt{\hbar}$ in order to properly decode the semiclassical structure of the quantum dynamics; see fig. \ref{fig:rmin_rmax} for visualization. Note that for $r<r_{\min}$, any calculation would become senseless, since these length scales are smaller than the initial uncertainty of the wavefunction. In any case, one must ensure that $|\Psi(0)\rangle$ is well reconstructed by the eigenbasis $|\phi_{n}\rangle$, namely, that $\sum_{n=1}^{N_{\text{states}}}|c_{n}|^{2}=1$. The equality might be replaced by an approximation sign, in order to speed up the numerical simulation, as long as a subset of the eigenstates contribute marginally to the sum (i.e., summing over their respective coefficients provide $\sum|c_{n}|^{2}\ll1$). One is free to define a threshold under which small $c_{n}$'s are omitted, i.e., their respective eigenstates are not counted throughout the calculation of $Q_{\infty}$. 

In order to evaluate the performance of QUPID across different systems and dynamical regimes, we first define a universal figure of merit. Throughout this section, we quantify the classical-quantum correspondence using a normalized difference parameter,
\begin{equation}
\label{normalized-difference}
\Delta\equiv\left|\frac{d_{c}-d_{q}}{d_{c}+d_{q}}\right|\cdot100\%,
\end{equation}
where we will use the superscripts $d_{q}^{(\text{g})}$ and $d_{q}^{(\text{MC})}$ to distinguish between grid-based QUPID and MC-based QUPID, respectively.

To extract these dimensions from the numerical data, we present log-log plots, where we rescale $r$ with respect to $r_{\min}$ and $\mathcal{N}(r)$ with respect to $\mathcal{N}(r_{\min})$, for both the classical and the quantum calculations. This rescaling emphasizes the slope of these linear curves, while the $y$-axis intercept is unnecessary for assessing the Pointwise Dimension. The slope is obtained using least-squares, in the range $0\leq\log(r/r_{\min})\leq\log(r_{\max}/r_{\min})$ where the exact upper bound is informed by the specific values of $\hbar$ and the dimensionality of the system.

Regarding the simulation parameters, classical trajectories are evolved using 4th order Runge-Kutta method (\texttt{ode45}) until $\tilde{T}=10^4$ (or $10^{6}$ iterations in the case of Kicked Rotor); for the relevant quantum simulations, the number of MC samples for the calculation of $Q_{n}$ is $5\cdot 10^4$; for all simulations we consider a set of reference points of size $M=100$. In most cases, we have considered $N_{x}=N_{\textbf{x}_{0}}=N_{\textbf{p}_{0}}=N_{\text{states}}$, where $N_{\textbf{x}}$ defines the real-space discretization and $N_{\textbf{x}_{0}},N_{\textbf{p}_{0}}$ define the discretization of the phase space gaussian basis. The parameters and numerical results are provided in the Appendix.

\subsection{$1D$ autonomous systems}
According to the Poincaré-Bendixson theorem,\citep{Lichtenberg1992RegularDynamics,Strogatz1994NonlinearChaos} in $1D$ autonomous systems, classical dynamics is strictly regular for all initial conditions, since energy is conserved; quantum dynamics should follow the same logic, so overall we expect $d_{c}\approx d_{q} \approx 1$. Although $1D$ systems are dynamically trivial, we find it helpful to examine QUPID in two cases to establish a baseline for the performance of the algorithm. In the first case, we consider the simple harmonic oscillator, $V(x)=\frac{1}{2}x^2$, and in the second case, we consider the quartic potential, $V(x)=\alpha x^{4}$, with $\alpha=0.05$.

\begin{figure}[ht!]
    \centering
    \includegraphics[width=1\linewidth]{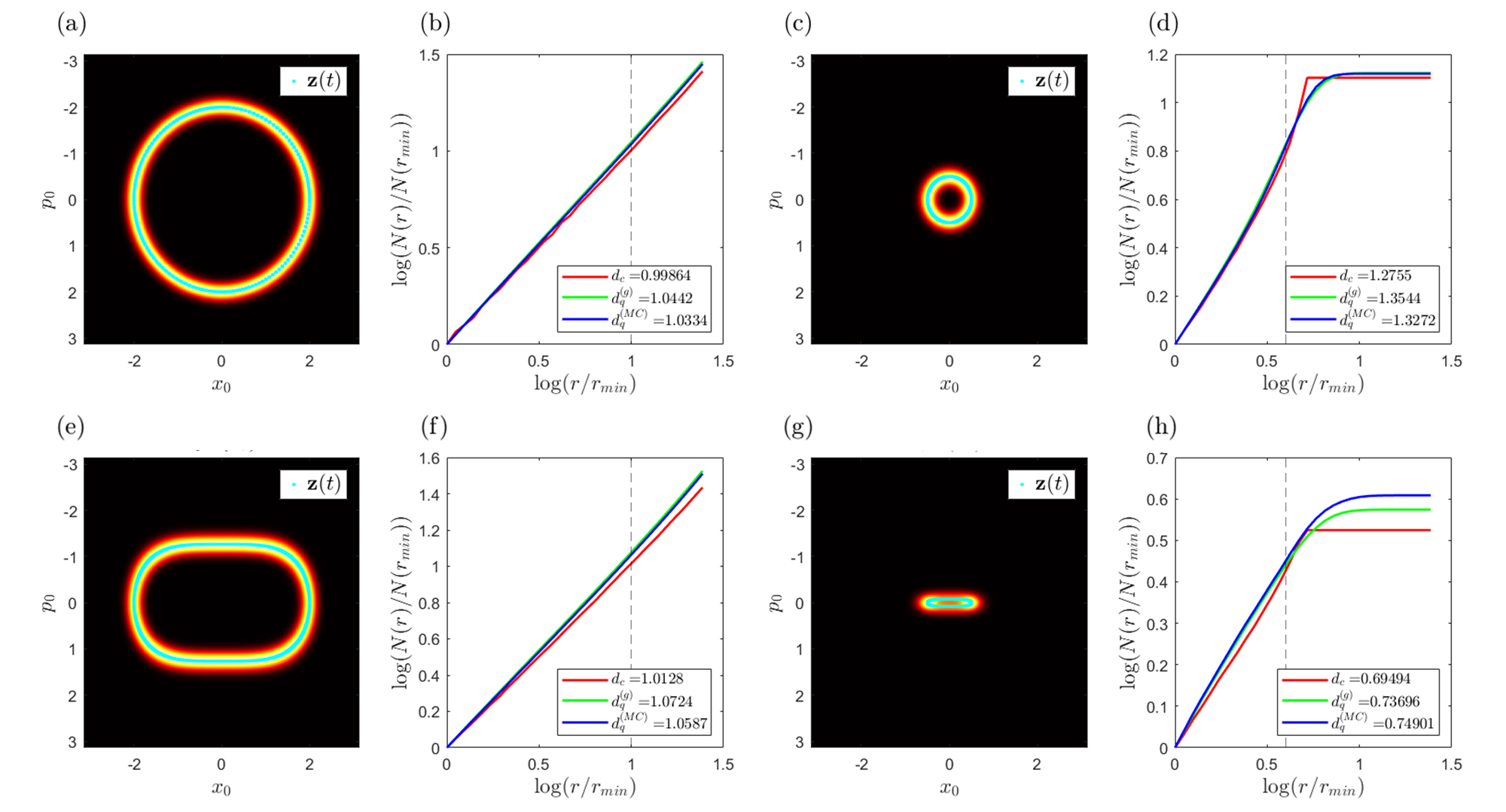}
\caption{Numerical evaluation of the Pointwise Dimension for $1D$ autonomous systems. Each pair of subfigures demonstrates $Q_{\infty}(\textbf{z}')$ overlapped by the classical trajectory $\textbf{z}(t)$, alongside the log-log plots yielding the Pointwise Dimension in the classical and quantum regimes. (a)-(d) are for the simple harmonic oscillator; (e)-(h) are for the quartic potential. (a)-(b) $\boldsymbol{\zeta}_{0}=(2,0)$ and (c)-(d) $\boldsymbol{\zeta}_{0}=(0.5,0)$; (e)-(f) $\boldsymbol{\zeta}_{0}=(2,0)$ and (g)-(h)  $\boldsymbol{\zeta}_{0}=(0.5,0)$. Note that, while in (b) and (f) the cutoff line for the computation of the slopes is at $\log r/r_{\min}=1$, we have selected $\log r/r_{\min}=0.6$ in (d) and (h).}
    \label{fig:qupid_1D}
\end{figure}

For both potentials, we consider a the initial condition $\boldsymbol{\zeta}_{0}=(2,0)$ without filtering out small contributions. To thoroughly assess the algorithm's convergence, we also consider an initial condition closer to the origin, $\boldsymbol{\zeta}_{0}=(0.5,0)$. Figure \ref{fig:qupid_1D} presents the classical and quantum Pointwise Dimensions obtained for both initial conditions using grid and MC. These converged results reflect the decreasing trend of $\Delta$ in Table \ref{tab:numerical_results_1D} as the sampling grid is refined. It is also evident from fig. \ref{fig:qupid_1D} that the classical trajectories perfectly overlap with $Q_{\infty}$.

The nonzero, yet small values of $\Delta$ mostly originate from numerical reasons: truncation of the Hilbert space, as depicted by finite $N_{\text{states}}$; and finite resolution of the spatial and coherent-state grids. The latter is reflected in the convergence of $\Delta$ as $\hbar$ decreases. However, these are just symptoms stemming from a profound reason. While the classical density is singular at the invariant torus, the quantum density is smoothed up to the scale of $\sqrt{\hbar}$, i.e., $d_{q}$ is sensitive to the exact value of $\hbar$. Excellent agreement ($\Delta\approx0$) can be approached only in the asymptotic limit $\hbar \to 0$, which demands an increasingly high resolution in the grid method. For this reason, one might consider the MC integration.

The importance of the integration method becomes particularly critical when $\boldsymbol{\zeta}_{0}$ is closer to the origin of phase space, such as $\boldsymbol{\zeta}_{0}=(0.5,0)$. In this regime, the torus structure of $Q_{\infty}$ is distorted and becomes more like a $2D$ spot, causing $\Delta$ to increase erroneously. While taking a smaller $\hbar$ helps resolve the torus, the MC method proves more robust in these sensitive regions. Essentially, MC integration provides a reliable path to capture the detailed Husimi structure by bypassing the accumulation of grid-based errors, enabling a proper treatment of all initial conditions. 

\subsection{$1D$ time-periodic system}
In contrast to $1D$ autonomous systems, $1D$ non-autonomous systems in classical dynamics can display either regular or chaotic motion, although energy is not conserved; quantum dynamics should follow the same logic, so overall we expect $d_{c}\approx d_{q} \approx 1$ in the regular case and $d_{c}\approx d_{q}\approx 2$ in the chaotic case. We examine QUPID for the exemplary case of Kicked Rotor $V(x)=K\cos x\sum_{n}(t-n)$ in three dynamical regimes. Recall that the classical Standard Map constitutes an analytic stroboscopic solution for Hamilton's equations with this potential:
\begin{equation}
\label{standard-map}
\begin{cases}
p_{n+1}=p_{n}+K\sin x_{n} \\
x_{n+1}=x_{n}+p_{n+1}
\end{cases}.
\end{equation}
In the first regime, with kick strength $K=0.1$, all trajectories are regular; in the second regime, with $K=1$, slightly beyond the limit where the last KAM torus breaks down, some trajectories are regular and some are chaotic; in the third regime, with $K=5$, chaos is evident in the vast majority of phase space. In each iteration, we consider $(x_{n},p_{n})\to\mod([x_{n},p_{n}],2\pi)-\pi$, which is allowed due to the cylindrical symmetry of the Standard Map. Also note that since $x_{n}$ is an angular coordinate, the distance between any two points in the cylindrical phase space is evaluated with a periodic metric,
\begin{equation}
\small
\label{periodic-metric}
D_{n}=\sqrt{(X_{n}-X(0))^{2}+(Y_{n}-Y(0))^{2}+(Z_{n}-Z(0))^{2}}=\sqrt{[\cos(x_{n})-\cos(x(0))]^2+[\sin(x_{n})-\sin(x(0))]^2+[p_{n}-p(0)]^2},
\end{equation}
where $X=\cos x,Y=\sin x$ and $Z=p$; eq. \ref{theta} applies similarly. 

\begin{figure}[ht!]
    \centering
    \includegraphics[width=1\linewidth]{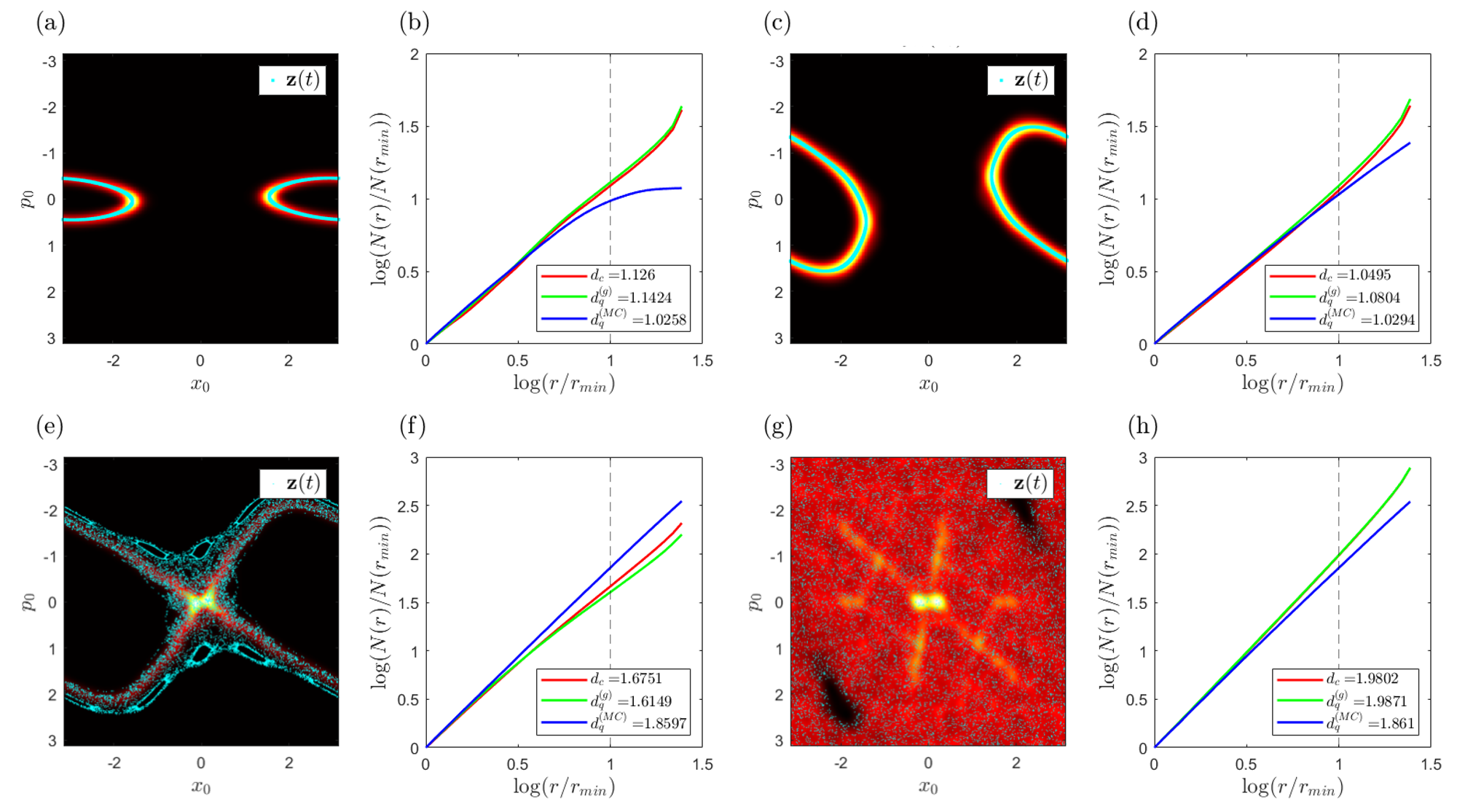}
\caption{Numerical evaluation of the Pointwise Dimension for the $1D$ non-autonomous, time-periodic system of the Kicked Rotor. Each pair of subfigures demonstrates $Q_{\infty}(\textbf{z}')$ overlapped by the classical trajectory $\textbf{z}(t)$, alongside the log-log plots yielding the Pointwise Dimension in the classical and quantum regimes. (a)-(b) $K=0.1$ with $\boldsymbol{\zeta}_{0}=(\pi/2,0)$; (c)-(d) $K=1$ with $\boldsymbol{\zeta}_{0}=(\pi/2,0)$, (e)-(f) $K=1$ with $\boldsymbol{\zeta}_{0}=(0.2,0)$; (g)-(h) $K=5$ with $\boldsymbol{\zeta}_{0}=(0.2,0)$. In subfigures (e) and (g), the classical trajectories were truncated in time, so they will not spoil the quantum distributions.}
    \label{fig:qupid_KR}
\end{figure}

For all kick strengths $K$, we have examined at the initial conditions $\boldsymbol{\zeta}_{0}=(\pi/2,0)$ and $\boldsymbol{\zeta}_{0}=(0.2,0)$, where each is employed in certain regimes, as detailed in Table \ref{tab:numerical_results_KR}. Figure \ref{fig:qupid_KR} presents the classical and quantum Pointwise Dimensions, obtained for both initial conditions using grid and MC which provide convergence in any case, as seen from the decreasing trend of $\Delta$ in Table \ref{tab:numerical_results_KR} as the sampling grid is refined. It is also evident from fig. \ref{fig:qupid_KR} that the classical trajectories perfectly overlap with $Q_{\infty}$; particularly, subfigure (f) is consistent with observation that in the semiclassical limit, the quantum localization of momentum almost vanishes.\citep{Delande2013KickedLocalization} 

The nonzero, yet small values of $\Delta$ originate from the same numerical reasons discussed above. The classical-quantum correspondence, as measured by $\Delta$, which is also observed in the log-log plots, is better in the chaotic cases, which do not exist in autonomous $1D$ systems. This is in contrast to the previous discussion on the $\hbar$-dependence in the regular case; in the ergodic case, the classical trajectory is already volume-filling, so the quantumness of $Q_{\infty}$ does not change its structure significantly compared to its classical counterpart. Here the grid method performs better than MC; however, the MC is more efficient for higher dimensions so we will focus on it in the $2D$ simulations.

\subsection{$2D$ autonomous systems}
In $2D$ autonomous systems, classical dynamics is either regular or chaotic, depending on the initial conditions; if quantum dynamics exhibits the same behavior as its classical counterpart, we anticipate $d_{c}\approx d_{q}\approx2$ in the regular case and $d_{c}\approx d_{q}\approx 3$ in the chaotic case. Before dealing with a coupled nonlinear system, let us analyze the simple harmonic oscillator again.

In the special case of a $2D$ isotropic simple harmonic oscillator, $V(x,y)=\frac{1}{2}(x^2+y^2)$, periodic dynamics is predicted for any initial condition, so $d_{c}\approx d_{q}\approx1$. This is a degenerate system, which seemingly challenges eq. \ref{Husimi-inf}. However, since the Hamiltonian in this case is separable, the eigenstates are product states, i.e., $|n_{x}\rangle\otimes|n_{y}\rangle$, and the Husimi representation depends only on the population of each energy subspace, being insensitive to off-diagonal terms within degenerate energy subspaces. Overall, after long-time averaging, the density operator describing the system becomes block-diagonal in energy, and eq. \ref{Husimi-inf} remains valid. We would find that the effective dimensionality remains the same as in $1D$. In contrast, for the anisotropic simple harmonic oscillator, we expect regular motion at higher effective dimensionality.

Now we consider the coupled quartic potential $V(x)=x^4+y^4+\lambda x^2y^2$ with $\lambda=4$. \citep{Caranicolas1987ChaosModel} Note that while its symmetry introduces spectral degeneracies, due to the interchangeability $x\leftrightarrow y $, the construction of $Q_{\infty}(\textbf{z}')$ remains valid because of the block-diagonal projection over degenerate subspaces. 

\begin{figure}[ht!]
    \centering
    \includegraphics[width=1\linewidth]{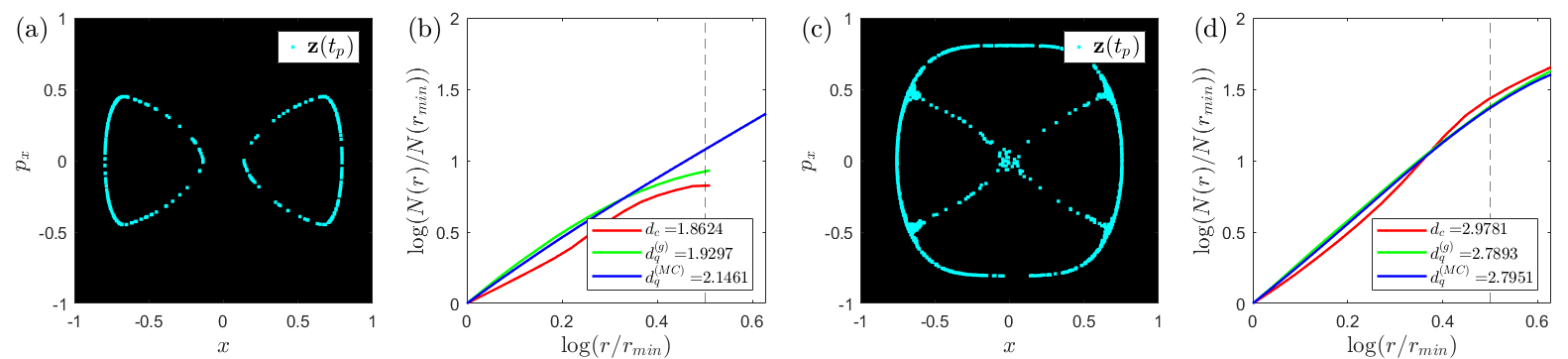}
\caption{Numerical evaluation of the Pointwise Dimension for a $2D$ autonomous system of a quartic coupled potential. Each pair of subfigures demonstrates the Poincaré surface of section of the classical trajectory $\textbf{z}(t)$, where $t_{p}$ are the recurrence times with $y=0$, alongside the log-log plots yielding the Pointwise Dimension in the classical and quantum regimes. (a)-(b) $\lambda=4$ with $\boldsymbol{\zeta}_{0}=(0.4,0.5,0.6,0)$, $E=0.43$ and (c)-(d) $\lambda=4$ with $\boldsymbol{\zeta}_{0}=(0.1,0.1,0.1,0.8)$, $E=0.33$. We have chosen not to show any section of $Q_{\infty}(\textbf{z}')$ as QUPID relies on full dimensional integration, which lower-dimensional slices cannot capture properly. Shown are the the MC results. Note that in (b) and (d) the cutoff line for the computation of the slopes is at $\log r/r_{\min}=0.5$. }
    \label{fig:qupid_2D}
\end{figure}

Classically, we obtain a mixed phase space, with chaos emerging mainly when the coupling term is dominant. We select representative initial conditions $\boldsymbol{\zeta}_{0}=(x(0),y(0),p_{x}(0),p_{y}(0))=(0.1,0.1,0.1,0.8)$, and $\boldsymbol{\zeta}_{0}=(0.4,0.5,0.6,0)$, which correspond to regularity and chaos, respectively. Figure \ref{fig:qupid_2D} presents their classical Poincaré surfaces of section, where these behaviors are evident, just as discussed in the context of fig. \ref{fig:Poincare}. Also presented are the classical and quantum Pointwise Dimensions, obtained for both initial conditions using the grid and MC. The MC calculations are converged, as seen from the decreasing trend of $\Delta$ in Table \ref{tab:numerical_results_KR} where the sampling grid is refined.

The nonzero, yet small values of $\Delta$ originate from the same numerical reasons discussed above. Note that these values are larger compared to the $1D$ time-periodic system, particularly in the regular case. These deviations might be attributed to quantum phenomena, e.g., tunneling, rather than grid resolution. The classical-quantum correspondence, as measured by $\Delta$, which is also observed in the log-log plots, is better in the chaotic case. In the regular case, the grid method performs much better than MC; in the chaotic case, they are comparable. However, the MC is more efficient.

\section{Can the Pointwise Dimension be extracted from the time autocorrelation function?}
The time autocorrelation of the wavepacket, $C(t)=\langle \phi(0)|\phi(t)\rangle$, is a measure of how closely the wavepacket at time $t$ overlaps the wavepacket at time $0$. The magnitude of the overlap $|C(t)|$ is therefore, loosely speaking, a measure of the distance between $\phi(t)$ and $\phi(0)$. This suggests that the height of the peaks in the autocorrelation function might correlate with the radii of the spheres that appear in the definition of the Pointwise Dimension. This conjecture will turn out to be precise for classical motion.

The connection between the classical autocorrelation function and the classical Pointwise Dimension leads to an alternative formulation of the latter that more closely resembles the quantum Pointwise Dimension, eqs. \ref{Husimi-inf}, \ref{dq-def}. This in turn gives us new tools to explore the connection between the quantum autocorrelation function $C(t)$ and the quantum Pointwise Dimension, QUPID. The latter connection is not analytical; however, guided by the classical equivalence we show that semi-quantitative reconstruction of the quantum Pointwise Dimension may be obtained from $C(t)$.

\subsection{Classical case}
\subsubsection{Reformulating the classical Pointwise Dimension as an integral over phase-space density}
The classical Pointwise Dimension was defined in Sec. II in terms of the frequency with which a trajectory revisits its initial phase-space coordinates $\textbf{z}_{0}$ as a function of the radius of a sphere in phase space centered at $\textbf{z}_{0}$. Here, we show that the classical Pointwise Dimension can be reformulated in terms of the density of classical distribution function. To do this, we associate the classical trajectory $\textbf{z}(t)$ with a frozen gaussian density,
\begin{equation}
\label{dis-1}
\rho_{\text{cs}}(\textbf{z}',\textbf{z}(t))=(2\pi\hbar)^{-N}\exp\left(-\frac{||\textbf{z}'-\textbf{z}(t)||^{2}}{2\hbar}\right)=(2\pi\hbar)^{-N}\exp\left(-\frac{D^{2}(t)}{2\hbar}\right).
\end{equation}
The corresponding integrated density within a sphere of radius $r$ centered around a reference point $\textbf{z}_{0}$ is
\begin{equation}
\label{dis-2}
\mathcal{N}_{c}(r,\textbf{z}(t);\textbf{z}_{0})=\int_{||\textbf{z}'-\textbf{z}_{0}||\leq r}\rho_{\text{cs}}(\textbf{z}',\textbf{z}(t))d\textbf{z}'.
\end{equation}
Since $\textbf{z}(t)$ is parametrized by $t$, we can write
\begin{equation}
\label{dis-3}
\mathcal{N}_{c}(r,t;\textbf{z}_{0})=\int_{||\textbf{z}'-\textbf{z}_{0}||\leq r}\rho_{\text{cs}}(\textbf{z}',t)d\textbf{z}'.
\end{equation}
This integral can be evaluated analytically: after rescaling the phase-space coordinates, the result is 
\begin{equation}
\label{dis-4}
\mathcal{N}_{c}(r,t,\textbf{z}_{0})=F_{\chi^{2}_{2N}}\left(\frac{r^{2}}{\hbar};\frac{D^{2}(t)}{\hbar}\right),
\end{equation}
where $\chi^{2}_{2N}$ is the non-central chi-squared distribution function, with $2N$ variables. $F_{\chi^{2}_{2N}}\left(\bar{r};\lambda\right)$ denotes the cumulative probability function associated with $\chi^{2}_{2N}$, loosely speaking the analog of a multidimensional error function. $\lambda$ denotes the instantaneous non-centrality parameter:
\begin{equation}
\label{dis-5}
\lambda(t)=\frac{||\textbf{z}(t)-\textbf{z}_{0}||^{2}}{\hbar}=\frac{D^{2}(t)}{\hbar}.
\end{equation}
\begin{figure}
    \centering
    \includegraphics[width=0.8\linewidth]{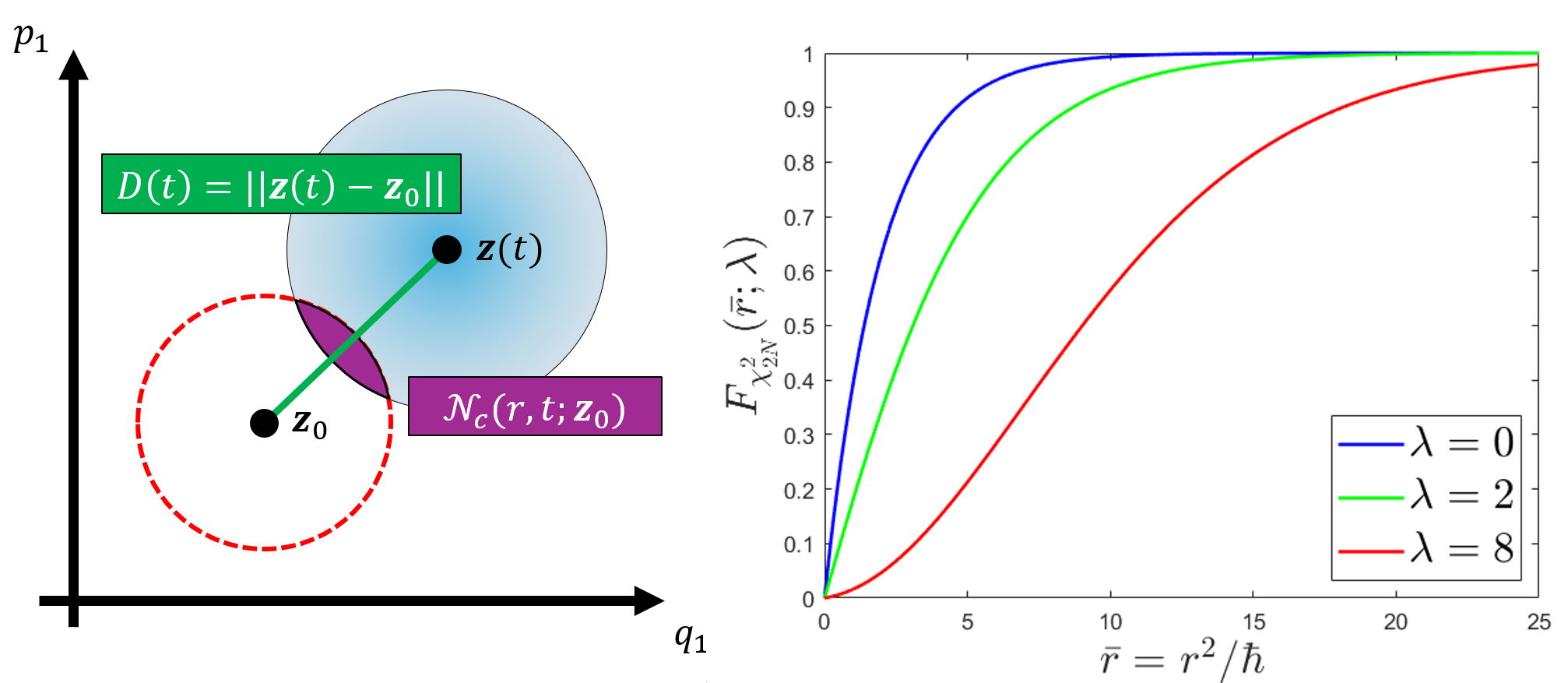}
    \caption{Left: the frozen gaussian centered at $\textbf{z}(t)$ (blue) and the integration sphere centered at $\textbf{z}_{0}$ (red). Right: after rescaling by $\sqrt{\hbar}$, the squared distance follows a non-central $\chi^{2}$ distribution with $2N$ dimensions and an instantaneous non-centrality parameter $\lambda(t)=D^{2}(t)/\hbar$. The cumulative distribution function $F_{\chi^{2}_{2N}}$ evaluated at $\bar r=r^{2}/\hbar$ gives the accumulated density $\mathcal{N}_{c}(r,t;\textbf{z}_{0})$.}
    \label{fig:chi}
\end{figure}
The time-average density entering the classical Pointwise Dimension is therefore
\begin{equation}
\label{dis-6}
\mathcal{N}_{c}(r;\textbf{z}_{0})=\lim_{\tilde{T}\to\infty}\frac{1}{\tilde{T}}\int_{0}^{\tilde{T}}\mathcal{N}_{c}(r,t;\textbf{z}_{0})dt=\lim_{\tilde{T}\to\infty}\frac{1}{\tilde{T}}\int_{0}^{\tilde{T}}F_{\chi^{2}_{2N}}\left(\frac{r^{2}}{\hbar};\frac{D^{2}(t)}{\hbar}\right)dt.
\end{equation}
Thus, the classical Pointwise Dimension has been reformulated in terms of a classical density $\rho_{\text{cs}}(\textbf{z}',\textbf{z}(t))$; the phase-space distance $D(t)$, which is at the core of the Martens-Ezra algorithm, enters only indirectly as the distance of the time-evolving coherent state from the reference point.

\subsubsection{The connection between the classical Pointwise Dimension and the cross-correlation functions $C^{\times}_{\text{cs}}(\textbf{z}',t)$ }
The classical Pointwise Dimension can be viewed as arising from a time-average of spatially integrated semiclassical cross-correlations. To make this mapping explicit, and connect to the quantum derivation, it is useful to define the classical coherent-state cross-correlation function,
\begin{equation}
\label{dis-7}
C^{\times}_{\text{cs}}(\textbf{z}',t)\equiv\langle\alpha_{\textbf{z}'}|\alpha_{\textbf{z}(t)}\rangle.
\end{equation}
In terms of density, $|C^{\times}_{\text{cs}}(\textbf{z}',t)|^{2}=\rho_{\text{cs}}(\textbf{z}',t)$, so eq. \ref{dis-2} can be re-written as the following spatial integral
\begin{equation}
\label{dis-8}
\mathcal{N}_{c}(r,t;\textbf{z}_{0})=(2\pi\hbar)^{-N}\int_{||\textbf{z}'-\textbf{z}_{0}||\leq r}
|C^{\times}_{\text{cs}}
(\textbf{z}',t)|^{2}d\textbf{z}'=(2\pi\hbar)^{-N}\int_{||\textbf{z}'-\textbf{z}_{0}||\leq r}|\langle\alpha_{\textbf{z}'}|\alpha_{\textbf{z}(t)}\rangle|^{2}d\textbf{z}'.\end{equation}
Equation \ref{dis-8} expresses the classical Pointwise Dimension as an integral over cross-correlation functions. 

\subsubsection{The classical Pointwise Dimension can be extracted directly from the autocorrelation function $C_{\text{cs}}(\textbf{z}',t)$}
The relation above (eq. \ref{dis-6}) is particularly useful because the phase-space distance $D(t)$ is encoded directly in the coherent-state autocorrelation function
\begin{equation}
\label{dis-9}
|C_{\text{cs}}(t)|^{2}=|\langle\alpha_{\textbf{z}_{0}}|\alpha_{\textbf{z}(t)}\rangle|^{2}=\exp\left(-\frac{D^{2}(t)}{2\hbar}\right)
\end{equation}
where $|\alpha_{\textbf{z}(t)}\rangle$ is a frozen gaussian following the classical trajectory $\textbf{z}(t)$, and therefore,
\begin{equation}
\label{dis-10}
\frac{D^{2}(t)}{2\hbar}=-\log|C_{\text{cs}}(t)|^{2}.
\end{equation}
Substituting eq. \ref{dis-10} into eq. \ref{dis-4} we obtain
\begin{equation}
\label{dis-10B}
\mathcal{N}_{c}(r,t;\textbf{z}_{0})=F_{\chi_{2N}^{2}}\left(\frac{r^{2}}{\hbar},-2\log|C_{\text{cs}}(t)|^{2}\right),
\end{equation}
and consequently
\begin{equation}
\label{dis-11}
\mathcal{N}_{c}(r;\textbf{z}_{0})=\lim_{\tilde{T}\to\infty}\frac{1}{\tilde{T}}\int_{0}^{\tilde{T}}\mathcal{N}_{c}(r,t;\textbf{z}_{0})dt=\lim_{\tilde{T}\to\infty}\frac{1}{\tilde{T}}\int_{0}^{\tilde{T}}F_{\chi^{2}_{2N}}\left(\frac{r^{2}}{\hbar};-2\log|C_{\text{cs}}(t)|^{2}\right)dt.
\end{equation}
Combining with eq. \ref{dis-6} and eq. \ref{Nrz-cl-av} establishes an explicit analytical mapping between the classical autocorrelation function $C_{\text{cs}}(t)$ and the classical Pointwise Dimension $d_{c}$:
\begin{equation}
\label{dis-12}
C_{\text{cs}}(t)\to\mathcal{N}_{c}(r,t;\textbf{z}_{0})\to\mathcal{N}_{c}(r;\textbf{z}_{0})\to d_{c}.
\end{equation}
Thus, using the frozen gaussian dynamics, the classical Pointwise Dimension can be obtained directly from the semiclassical autocorrelation function without explicitly reconstructing the trajectory in phase space.

Comparing eq. \ref{dis-11} with eq. \ref{dis-8}, which expresses $\mathcal{N}_{c}(r,t;\textbf{z}_{0})$ in terms of an integral over cross-correlation functions, leads to the surprising conclusion that a time-series can encode spatial information. This paradox can be reconciled by noting that the gaussian structure of the coherent states reduces the entire spatial dependence to a single coordinate $r$. 

\subsection{Quantum case}
\subsubsection{Motivation}
In Section III, we gave a definition of a quantum Pointwise Dimension, $d_q,$ and we showed that it gave semi-quantitative correspondence with the classical Pointwise Dimension, $d_c$. In particular, we showed that this definition of $d_q$ can clearly distinguish between quantum systems characterized by regular vs chaotic motion in the classical regime.

Our definition of $d_q$, based on eqs. \ref{Husimi-inf}, and \ref{Nrz-q} involves: (1) a spatial integral over wavepacket cross-correlation functions and (2) a time integral. But in the classical case, we found that, surprisingly, the spatial integral, eq. \ref{dis-8}, can be obtained from just the wavepacket autocorrelation function, eq. \ref{dis-11}. This raises the question whether the same might be possible in the quantum case. Admittedly, the classical case is special in that there is a closed form expression for the wavepacket dynamics $|\Psi(t)\rangle=|\alpha_{\textbf{z}(t)}\rangle $ and $\rho_{\text{cs}}=|\alpha_{\textbf{z}(t)}|^2$. However, from an informational point of view, if in the classical regime the integral over cross-correlation functions can be obtained just from the autocorrelation function, it seems plausible that the same would hold in the quantum regime even though there is no closed form expression for the time-evolving wavepacket.

To develop this line of thinking further, consider the quantum mapping analogous to the classical mapping, eq. \ref{dis-12}:
\begin{equation}
\label{dis-13B}
C(t) \to \mathcal{N}_{q}(r,t;\textbf{z}_0) \to \mathcal{N}_{q}(r,\textbf{z}_0) \to d_q
\end{equation}
The unknown step is the first one, $C(t)\to \mathcal{N}_{q}(r,t;\textbf{z}_{0})$; once $ \mathcal{N}_{q}(r,t;\textbf{z}_{0})$ is known, the last two steps are just integrations over $t$ and over $r$. Despite the uncertainty in the first step, there are certain considerations that mitigate this uncertainty:

(i) We know from Section IV that $d_q$ is in qualitative agreement with $d_c$. This suggests that even if $\mathcal{N}_q $ is very different from $\mathcal{N}_c$, the mapping  $\mathcal{N}_q(r,t;\textbf{z}_{0})\to \mathcal{N}_q(r,\textbf{z}_{0})\to d_q$ is robust, in the sense that significant differences between $\mathcal{N}_q$ and $\mathcal{N}_c$ may wash out when integrating over $r$ and $t$. As a result, we do not have to do a perfect job of reconstructing $\mathcal{N}_q(r,t;\textbf{z}_{0})$ from $C(t)$; as long as the qualitative behavior of $\mathcal{N}_{q}(r,t;\textbf{z}_{0})$ is correct, the error will to some extent cancel out. 

(ii) Note that $C(t)$ is a special case of $C^{\times}(t)$, where $|\alpha_{\textbf{z}’}\rangle=|\alpha_{\textbf{z}_{0}}\rangle$. In the limit of small $r$,
\begin{equation}
C(t)=\lim_{r\to0}C^{\times}(r,t)
\end{equation}
and therefore is expected dominate the integral $\int|C^{\times}(\textbf{z}’,t)|^2 d\textbf{z}’$ in eq. \ref{dis-8}. This suggests that at least in the small $r$ region, we should be able to recover $\mathcal N_{q}(r,t;\textbf{z}_0)$ accurately from $C(t)$. 

\begin{figure}[ht!]
    \centering
    \includegraphics[width=0.8\linewidth]{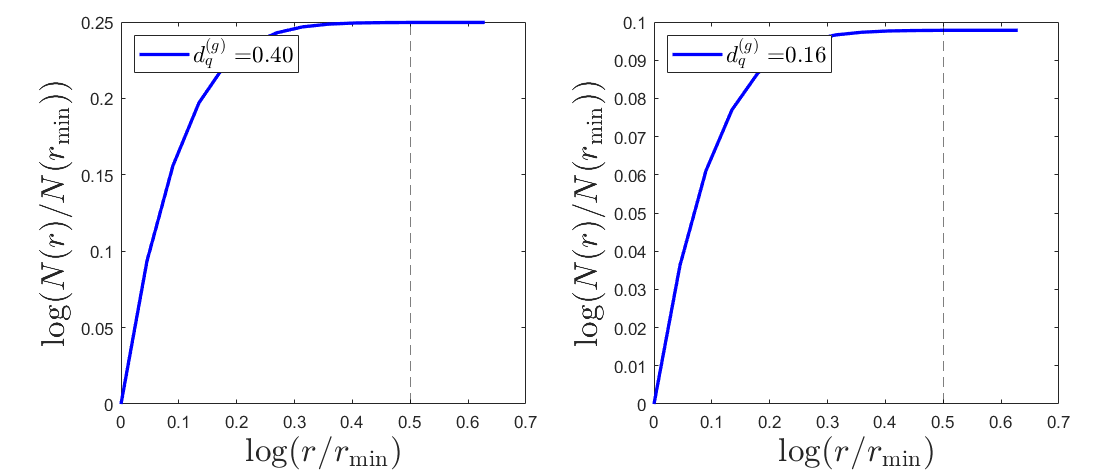}
\caption{A naive attempt to extract the quantum Pointwise Dimension via the quantum autocorrelation function $C(t)$. Shown are the results for the regular motion (left) and the chaotic motion (right) in the $2D$ coupled quartic potential of Sec. IV. C.}
    \label{fig:AC_pointwise}
\end{figure}

The simplest strategy to extract the quantum Pointwise Dimension from the quantum autocorrelation function would be to construct an analog of the classical procedure which assigns a distance to a height in the correlation function. Essentially, this is just eq. \ref{dis-11} with $\mathcal{N}_{c}(r,\textbf{z}_{0})$ replaced by $\mathcal{N}_{q}(r,\textbf{z}_{0})$ and $C_{\text{cs}}(t)$ replaced by $C(t)$. Figure \ref{fig:AC_pointwise} shows the results for the $2D$ quartic potential of Sec. IV. C., where the strategy is seem to fail completely: for regular motion, $d_{q}$ is $0.4$ instead of $2$ and for chaotic motion, $d_{q}$ is $0.16$ instead of $3$. Despite this failure, we conjecture that the wavepacket recurrences in $C(t)$ still contain qualitative information about the proximity of $\phi(t)$ to $\phi(0)$ in phase space. An alternative approach to extracting this information is described in the next subsection.

\textit{Machine-learning prediction of quantum Pointwise Dimension from classical autocorrelation functions.} To examine if $C(t)$ can indeed distinguish regular from chaotic motion, we adopt a machine learning approach. The strategy is to train a machine-learning algorithm on the classical hierarchy
\begin{equation}
C_{\text{cs}}(t)\to\mathcal{N}_{c}(r,t;\textbf{z}_{0})\to\mathcal{N}_{c}(r;\textbf{z}_{0})\to d_{c}
\end{equation}
to achieve the quantum analog
\begin{equation}
C(t)\to\mathcal{N}_{q}(r,t;\textbf{z}_{0})\to\mathcal{N}_{q}(r;\textbf{z}_{0})\to d_{q}.
\end{equation}
Specifically, the approach is to learn the classical mapping from $C_{\text{cs}}(t)\to\mathcal{N}_{c}(r,t;\textbf{z}_{0})$ without informing the learning algorithm about the underlying distances $D(t)$ and then use the training to map $C(t)\to\mathcal{N}_{q}(r,t;\textbf{z}_{0})$.

\begin{figure}[ht!]
    \centering
    \includegraphics[width=0.6\linewidth]{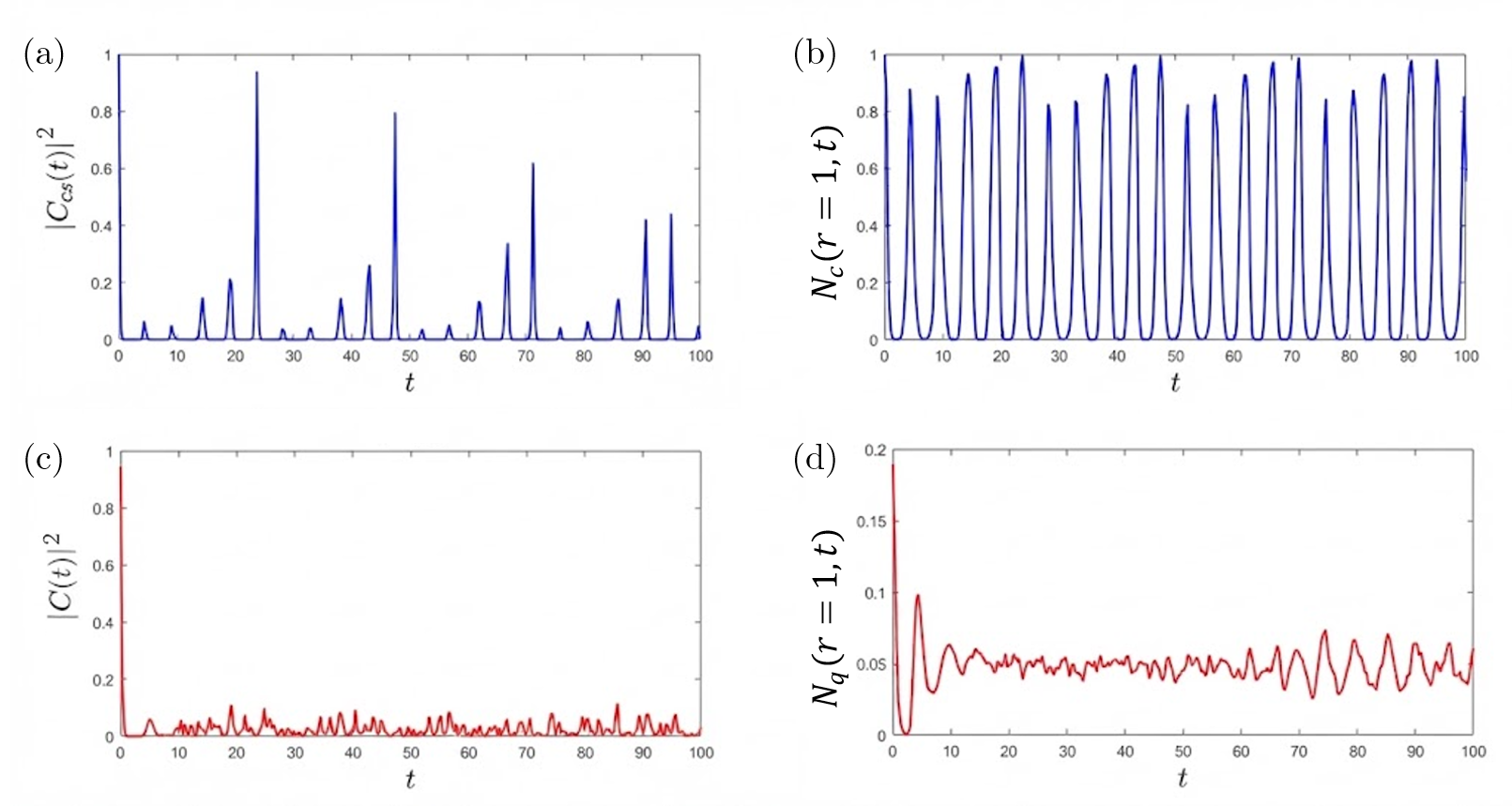}
  \caption{The passage from (a) $C_{\text{cs}}(t)$ to (b) $\mathcal{N}_{c}(r,t;\textbf{z}_{0})$ for the case of classically regular dynamics. The passage from (c) $C(t)$ to (d) $\mathcal{N}_{q}(r,t;\textbf{z}_{0})$ for the corresponding quantum case. The strategy will be to train a machine-learning algorithm on the classical passage and then give it the quantum correlation function $C(t)$ and ask it to find $\mathcal{N}_{q}(r,t;\textbf{z}_{0})$. The results shown here are for the $2D$ coupled quartic potential, see text.}
    \label{fig:regular}
\end{figure}

\begin{figure}[ht!]
    \centering
    \includegraphics[width=0.6\linewidth]{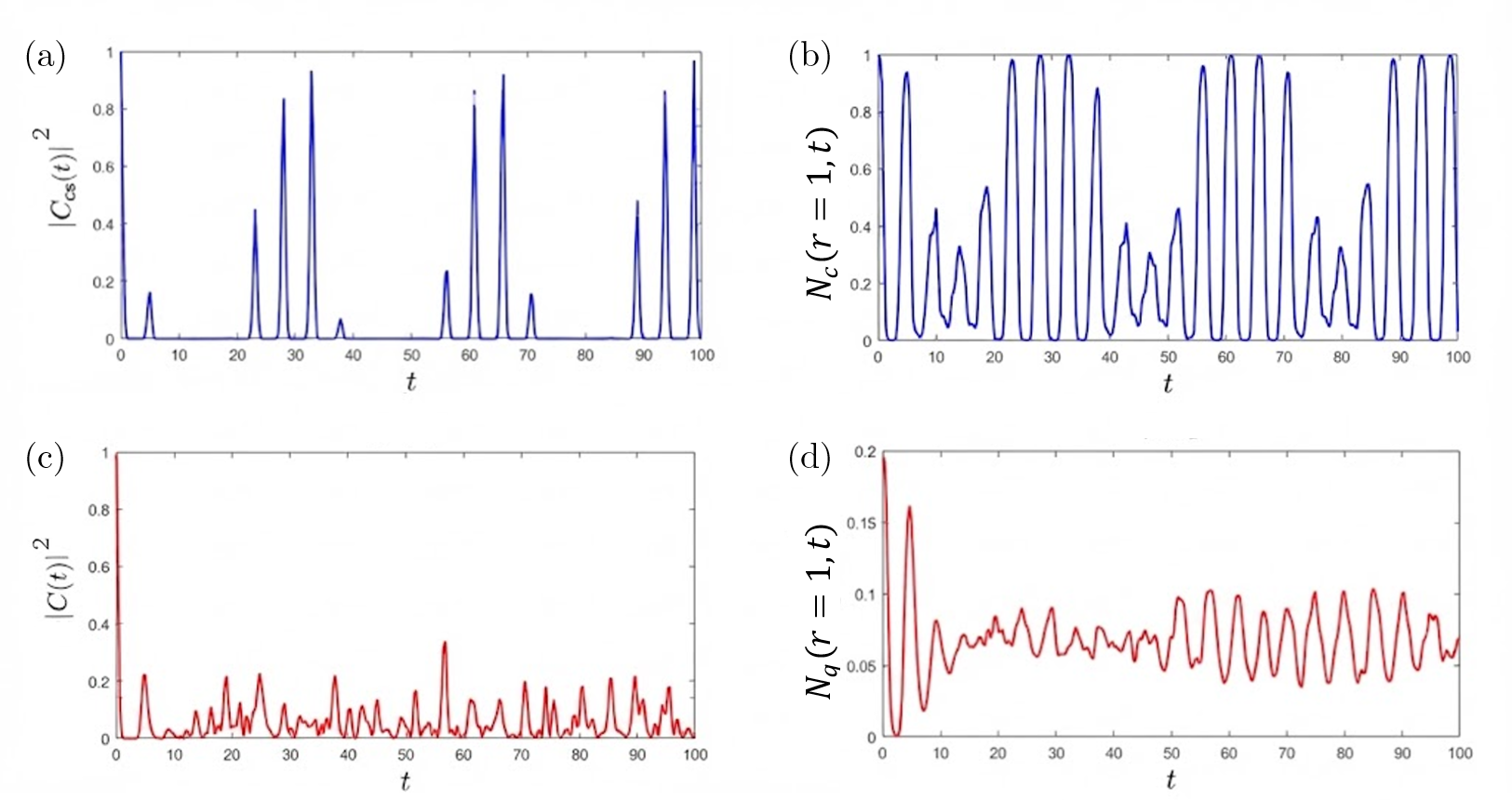}
    \caption{The passage from (a) $C_{\text{cs}}(t)$ to (b) $\mathcal{N}_{c}(r,t;\textbf{z}_{0})$ for chaotic dynamics. The passage from (c) $C(t)$ to (d) $\mathcal{N}_{q}(r,t;\textbf{z}_{0})$ for chaotic dynamics. Note that the classical passage is completely analytical. The strategy will be to train a machine-learning algorithm on the classical passage and then give it the quantum correlation function $C(t)$ and ask it to find $\mathcal{N}_{q}(r,t;\textbf{z}_{0})$. The results shown here are for the $2D$ coupled quartic potential, see text.}
    \label{fig:chaotic}
\end{figure}

Figure \ref{fig:regular} (a) and (b) depict the passage from $C_{\text{cs}}(t)$ to $\mathcal{N}_{c}(r,t)$ in the case of classically regular motion and (c) and (d) depict the passage from $C(t)$ to $\mathcal{N}_{q}(r,t)$ in the corresponding quantum case, for the $2D$ coupled quartic potential of Sec. IV. C. Figure \ref{fig:chaotic} depicts the passage from $C_{\text{cs}}(t)$ to $\mathcal{N}_{c}(r,t)$ and $C(t)$ to $\mathcal{N}_{q}(r,t)$ in the classically chaotic case. In the classical case, both for regular and chaotic motion, the passage from $C_{\text{cs}}(t)$ to $\mathcal{N}_{c}(r,t)$ is analytical. The strategy of the machine learning is to train on the classical case and then to pass from $C(t)$ to $\mathcal{N}_{q}(r,t)$ in the quantum case, where the passage is not analytical.

\begin{figure}[ht!]
    \centering
    \includegraphics[width=0.7\linewidth]{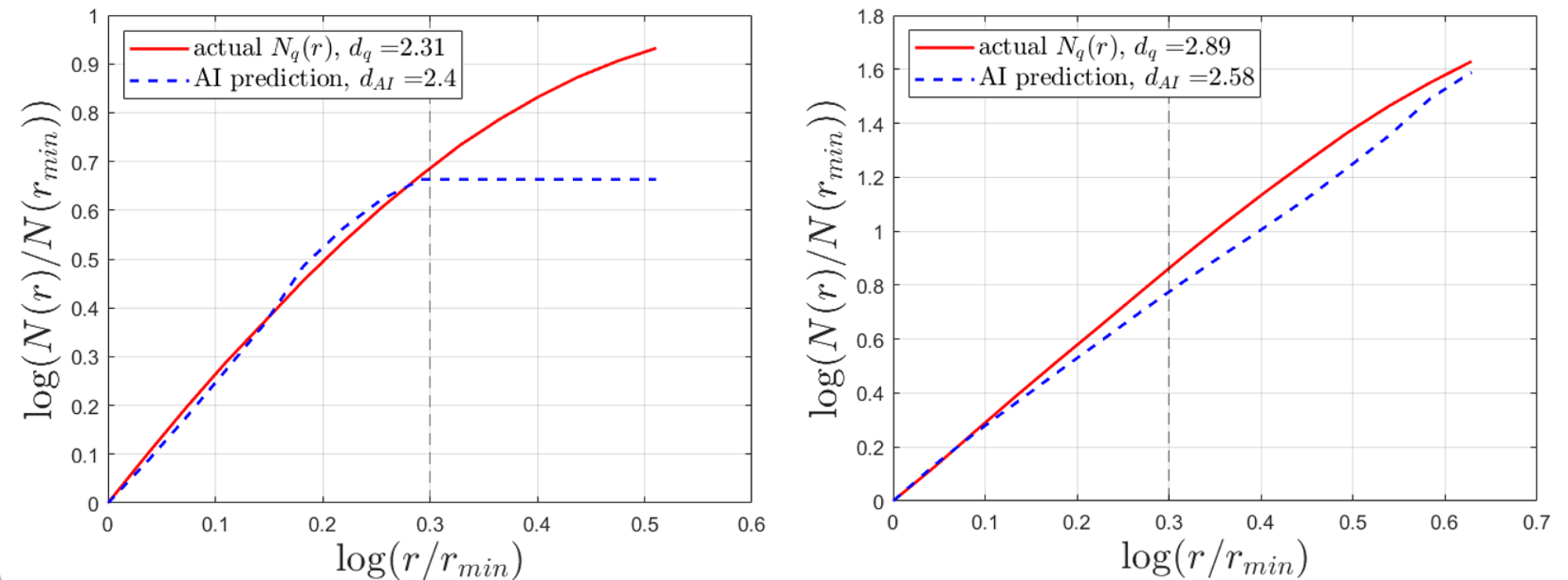}
    \caption{Comparison between the exact quantum Pointwise Dimension and the classical trajectory based machine learning predictions for the $2D$ coupled quartic potential. The machine-learning model, here denoted as "AI", leads to promising predictions within the fitting range, both in the regular case (left) and in the chaotic case (right). }
    \label{fig:ML_predictions}
\end{figure}

The learning algorithm is based on a collection of classical initial conditions integrated via Hamilton's equations of motion. For each classical trajectory, the algorithm records the autocorrelation function $C_{\text{cs}}(t)$ and the classical density $\mathcal{N}_{c}(r,t)$. The inference step is based on a supervised 1-nearest neighbor matching:\citep{Cover1967NearestClassification} for a given quantum autocorrelation function $C(t)$, the algorithm minimizes the quantity $\int|C_{\text{cs}}(t)-C(t)|^{2}dt$ across the collection of classical trajectories and selects the $\mathcal{N}_{c}(r,t)$ originating from the best-matching classical trajectory as the predicted quantum density $\mathcal{N}_{q}(r,t)$. Figure \ref{fig:ML_predictions} shows the promising results of this training scheme, yielding excellent agreement with the exact results from the quantum calculations, based on the Husimi representation. 

The occurrence of a plateau in the regular case originates from the torus structure of regular trajectories; at large radii, the classical density vanishes, but a quantum density can still exist there. Since the nearest-neighbors algorithm relies strictly on the classical trajectories, it cannot reproduce the quantum behavior beyond the classical boundary. For this reason, we have chosen to set the cutoff line for the computation of the slopes at $\log r/r_{\min}=0.3$ in the regular case, but also in the chaotic case for the sake of consistency.

\subsection{$\mathcal{N}_{q}(r,t;\textbf{z}_{0})$ vs $P(a|a)$: The loss of information by integrating over time}
Having shown that the Pointwise Dimension can be constructed from the autocorrelation function, we now ask what is lost when $C(t)$ is integrated over time. We focus here on the quantum case, although similar arguments apply to the classical case. We begin by discussing $P(a|a)$, which is a physically appealing metric of the time a quantum state spends in the vicinity of where it started.\citep{Nordholm1974QuantumMolecules,Heller1987QuantumSpace,Heller2018TheSpectroscopy}

$P(a|a)$ can be obtained from the time autocorrelation function,
\begin{equation}
\label{Paa-rel}
P(a|a)=\lim_{\tilde{T}\to\infty}{\frac{1}{\tilde{T}}\int_{0}^{\tilde{T}}|C(t)|^2}dt=\sum_{n=1}^{N_{\text{states}}}|c_{n}|^{4}=\frac{1}{ N_{\text{states}}}\int_{-\infty}^{\infty}|Q_{\infty}(\textbf{z}')|^{2}d\textbf{z}'=\frac{1}{N_{\infty}},
\end{equation}
where the second equality is correct if the system is non-degenerate. $P(a|a)$ is often called the Inverse Participation Ratio (IPR), 
\begin{equation}
P(a|a)=\text{IPR},
\end{equation}
which measures the accessed phase-space volume $N_{\infty}$. \citep{Heller1987QuantumSpace} IPR successfully distinguishes localization ($\text{IPR}\approx1$) vs. ergodicity ($\text{IPR}\sim1/N_{\text{states}}$). The main difference between these volume-based metrics and QUPID is that they reflect the overall exploration of phase space regardless of its dimension, while QUPID evaluates the effective dimensionality of quantum dynamics.

Note that $P(a|a)$ cannot distinguish two systems which accumulate comparable total recurrence weight near the initial point while distributing that weight very differently in time and in radius. As an example, consider a $3D$ anisotropic harmonic oscillator with rationally independent frequencies. A classical trajectory will live on a $3D$ torus. In the quantum case, let us start with a coherent state $|\Psi(0)\rangle=|\alpha_{x}\rangle\otimes|\alpha_{y}\rangle\otimes|\alpha_{z}\rangle$ such that $|\alpha_{x}|^{2}=|\alpha_{y}|^{2}=|\alpha_{z}|^2=\alpha_{0}$.  In this case, one can show that $P^{(1)}(a|a)=[e^{-2\alpha_{0}}I_{0}(2\alpha_{0})]^{3}$, where $I_{0}$ is the modified Bessel function of the first kind. For the specific choice $\alpha_{0}\approx0.362$, we obtain that $P^{(1)}(a|a)=1/6$.

On the other hand, consider any $2D$ non-degenerate chaotic system (e.g., asymmetric chaotic billiard), starting with $|\Psi(0)\rangle=\frac{1}{6}\sum_{n=1}^{6}|E_{n}\rangle$ where $|E_{n}\rangle$ is an eigenstate. We find that $P^{(2)}(a|a)=\sum_{n=1}^{4}|c_{n}|^{4}=1/6$,  so that $P^{(1)}(a|a)=P^{(2)}(a|a)$. This example shows the loss of information upon time-averaging.

Finally, note that the explicit connection between these $P(a|a)$ and $\mathcal{N}_{q}(r,t;\textbf{z}_{0})$ is the following:
\begin{equation}
\label{dis-15}
P(a|a)=Q_{\infty}(\textbf{z}_{0})=\lim_{r\to0}\frac{\mathcal{N}_{q}(r,\textbf{z}_{0})}{V_{2N}(r)}=\lim_{r\to0}\frac{1}{V_{2N}(r)}\left[\lim_{\tilde{T}\to\infty}\frac{1}{\tilde{T}}\int_{0}^{\tilde T}\mathcal{N}_{q}(r,t;\textbf{z}_{0})dt\right],
\end{equation}
where $V_{2N}(r)$ is the volume of a sphere of radius $r$ in the $2N$-dimensional phase space. Equation \ref{dis-15} shows that $P(a|a)$ is a time-averaged measure, whereas $\mathcal{N}(r,t;\textbf{z}_{0})$ retains the temporal and spatial information.

\section{Conclusion}
In this work, we introduced the Quantum Pointwise Dimension (QUPID) as a quantitative diagnostic of quantum chaos in moving wavefunctions. While most previous works have focused either on the morphology of eigenstates or on spectral statistics, QUPID constitutes a natural extension of the Pointwise Dimension, a tool that has proved useful as a diagnostic of chaos in classical mechanics. We have found this extension distinguishes between regularity and ergodicity in the quantum regime, in good correspondence with the classical dynamics.

By using the Husimi representation, QUPID extracts the effective dimensionality of the underlying classical dynamics without computationally expensive real-time wavepacket propagation. We have demonstrated excellent classical-quantum correspondence in the semiclassical limit in $1D$ autonomous systems, in a $1D$ time-periodic system, and in a $2D$ autonomous system with mixed phase space. The method appears robust at distinguishing between confinement to invariant tori and ergodic motion on energy surfaces.

We reformulated the classical and quantum Pointwise Dimension in a way that highlighted their similarity. In the process, we showed that the classical Pointwise Dimension can be extracted rigorously from the classical time autocorrelation function, while the quantum Pointwise Dimension can be extracted semi-quantitatively from the quantum time autocorrelation function using machine learning trained on the classical case.

Several further directions come to mind. The most natural extension is to apply QUPID to high-energy eigenstates, which historically have been central to the study of quantum chaos; QUPID could be combined with other diagnostics to quantify the effective dimensionality of scarred states. Furthermore, it might be interesting to extend the QUPID formalism from pure states to mixed states. 

\section{Acknowledgements}
J. B. thanks Jacob Higer and Shaked Levy for helpful discussions. Financial support for this work was provided by the German-Israeli Foundation for Scientific Research and Development (GIF), the Israel Science Foundation (Grant Nos. 1094/16 and 1404/21), the Minerva Foundation, and the historic generosity of the Harold Perlman family. Figures \ref{fig:pointwise_artwork} and \ref{fig:alexander} were generated using \textit{Google Gemini 3} for illustrative purposes to represent the concepts of the Pointwise Dimension and Monte-Carlo methods, respectively.  The authors have verified the scientific accuracy of all elements within the figures and affirm that no original research data was manipulated or generated by AI tools. 

\section{Appendix: Detailed numerical data}

For the $1D$ simulations, in both the simple harmonic oscillator and quartic potential, we initially considered a "sparse" and a "dense" parameter set (see Table \ref{tab:numerical_results_1D}) without filtering out small contributions. For the time-periodic $1D$ simulations, for all all kick strengths $K$, we initially considered a "sparse" and a "dense" parameter set (see Table \ref{tab:numerical_results_KR}) without filtering out small contributions. For the $2D$ simulations, we initially consider a "sparse" and a "dense" parameter set (see Table \ref{tab:numerical_results_2D}), but in the end applied this only for the MC integration; for the grid, we select only one set of parameters, anticipating the departure from grid-based calculations at higher dimensions. In the grid calculation we have thrown away about $50\%$ of the eigenstates owing to their negligible contributions (with the threshold of $|c_{n}|^{2}\geq10^{-6}$).

\begin{table}[ht!]
\centering
\small
\renewcommand{\arraystretch}{1.3}
\begin{tabular}{|c|c|c|c|c||c|c|c|c|}
\hline
Potential & \multicolumn{4}{c||}{Simple harmonic oscillator} & \multicolumn{4}{c|}{Quartic, $\alpha=0.05$} \\ \hline
Method & \multicolumn{2}{c|}{Grid} & \multicolumn{2}{c||}{MC} & \multicolumn{2}{c|}{Grid} & \multicolumn{2}{c|}{MC} \\ \hline
Resolution & \textit{Sparse} & \textit{Dense} & \textit{Sparse} & \textit{Dense} & \textit{Sparse} & \textit{Dense} & \textit{Sparse} & \textit{Dense} \\ \hline
$\boldsymbol{\zeta}_{0}$ & \multicolumn{4}{c||}{$(2,0)$} & \multicolumn{4}{c|}{$(2,0)$} \\ \hline
$\hbar$ & $\pi/100$ & $\pi/200$ & $\pi/100$ & $\pi/200$ & $\pi/100$ & $\pi/200$ & $\pi/100$ & $\pi/200$ \\ \hline
$N_{\text{states}}$ & 100 & 200 & 100 & 200 & 100 & 200 & 100 & 200 \\ \hline
$d_{c}$ & \multicolumn{4}{c||}{0.9986} & \multicolumn{4}{c|}{    1.0128} \\ \hline
$d_{q}$ &     1.0787& 1.0442&     1.0560&     1.0334&     1.1169&     1.0724&     1.0854&     1.0587\\ \hline
$\Delta$ (\%) & 3.9& 2.2& 2.8& 1.7& 4.9& 2.9& 3.5& 2.2\\ \hline \hline 
$\boldsymbol{\zeta}_{0}$ & \multicolumn{4}{c||}{$(0.5,0)$} & \multicolumn{4}{c|}{$(0.5,0)$} \\ \hline
$\hbar$ & $\pi/100$ & $\pi/200$ & $\pi/100$ & $\pi/200$ & $\pi/100$ & $\pi/200$ & $\pi/100$ & $\pi/200$ \\ \hline
$N_{\text{states}}$ & 100 & 200 & 100 & 200 & 100 & 200 & 100 & 200 \\ \hline
$d_{c}$ & \multicolumn{4}{c||}{    1.2755} & \multicolumn{4}{c|}{0.6949} \\ \hline
$d_{q}$ &     1.3915& 1.3544& 1.3605& 1.3272& 0.8044& 0.7370& 0.8484 & 0.7490\\ \hline
$\Delta$ (\%) & 4.4& 3& 3.2& 2& 7.3& 2.9& 10& 3.7\\ \hline
\end{tabular}
\vspace{10pt}
\caption{Numerical parameters and convergence results for $1D$ systems, grouped by the potential and initial conditions.}
\label{tab:numerical_results_1D}
\end{table}

\begin{table}[ht!]
\centering
\small
\renewcommand{\arraystretch}{1.3}
\begin{tabular}{|c|c|c|c|c||c|c|c|c|}
\hline
$K$& \multicolumn{4}{c||}{0.1} & \multicolumn{4}{c|}{1} \\ \hline
Method & \multicolumn{2}{c|}{Grid} & \multicolumn{2}{c||}{MC} & \multicolumn{2}{c|}{Grid} & \multicolumn{2}{c|}{MC} \\ \hline
Resolution & \textit{Sparse} & \textit{Dense} & \textit{Sparse} & \textit{Dense} & \textit{Sparse} & \textit{Dense} & \textit{Sparse} & \textit{Dense} \\ \hline
$\boldsymbol{\zeta}_{0}$ & \multicolumn{4}{c||}{$(\pi/2,0)$} & \multicolumn{4}{c|}{$(\pi/2,0)$} \\ \hline
$\hbar$ & $\pi/100$ & $\pi/200$ & $\pi/100$ & $\pi/200$ & $\pi/100$ & $\pi/200$ & $\pi/100$ & $\pi/200$ \\ \hline
$N_{\text{states}}$ & 200& 400& 200& 400& 200& 400& 200& 400\\ \hline
$d_{c}$ & \multicolumn{4}{c||}{1.1260} & \multicolumn{4}{c|}{ 1.0495} \\ \hline
$d_{q}$ & 1.1730& 1.1424& 1.1496& 1.0258& 1.1198& 1.0804& 1.0884& 1.0294\\ \hline
$\Delta$ (\%) & 2& 0.7& 1& 4.7& 3.2& 1.5& 1.8& 1\\ \hline \hline 
$K$& \multicolumn{4}{c||}{1} & \multicolumn{4}{c|}{5} \\\hline
Method & \multicolumn{2}{c|}{Grid}& \multicolumn{2}{c||}{MC}& \multicolumn{2}{c|}{Grid}& \multicolumn{2}{c|}{MC}\\\hline
Resolution & \textit{Sparse} & \textit{Dense} & \textit{Sparse} & \textit{Dense} & \textit{Sparse} & \textit{Dense} & \textit{Sparse} &\textit{Dense} \\\hline
$\boldsymbol{\zeta}_{0}$ & \multicolumn{4}{c||}{$(0.2,0)$}& \multicolumn{4}{c|}{$(0.2,0)$}\\\hline 
$\hbar$ & $\pi/100$ & $\pi/200$ & $\pi/200$& $\pi/400$& $\pi/100$ & $\pi/200$ & $\pi/100$ & $\pi/200$ \\ \hline
$N_{\text{states}}$ & 200& 400& 400& 400& 200& 400& 200& 400\\ \hline
$d_{c}$ & \multicolumn{4}{c||}{1.6751} & \multicolumn{4}{c|}{1.9802} \\ \hline
$d_{q}$ & 1.6314& 1.6149& 1.8597& 1.7545& 1.9885& 1.9871& 1.8972& 1.8610\\ \hline
$\Delta$ (\%) & 1.3& 1.8& 5.2& 2.3& 0.2& 0.2& 2.1& 3.1\\ \hline
\end{tabular}
\vspace{10pt}
\caption{Numerical parameters and convergence results for the Kicked Rotor, grouped by the kick strength $K$ and initial conditions.}
\label{tab:numerical_results_KR}
\end{table}

\begin{table}[ht!]
\centering
\small
\renewcommand{\arraystretch}{1.3}
\begin{tabular}{|c|c|c|c||c|c|c|}
\hline
$\lambda$& \multicolumn{3}{c||}{4}& \multicolumn{3}{c|}{4}\\\hline
Method & Grid & \multicolumn{2}{c||}{MC} & Grid & \multicolumn{2}{c|}{MC}\\\hline
Resolution & --- & \textit{Sparse} & \textit{Dense} & --- & \textit{Sparse} & \textit{Dense} \\\hline
$\boldsymbol{\zeta}_{0}$ & \multicolumn{3}{c||}{$(0.4,0.5,0.6,0)$} & \multicolumn{3}{c|}{$(0.1,0.1,0.1,0.8)$}\\\hline 
$\hbar$ & $\pi/50$ & $\pi/50$ & $\pi/100$ & $\pi/50$ & $\pi/50$ & $\pi/100$\\ \hline
$N_{\text{states}}$ $\left[N_{\text{states}}^{\text{(eff)}}\right]$& 200 [105] & 100 & 100 & 100 [49] & 100 & 200\\ \hline
$d_{c}$ & \multicolumn{3}{c||}{1.8624} & \multicolumn{3}{c|}{2.9781}\\ \hline
$d_{q}$ & 1.9297 & 2.2399 & 2.1461 & 2.7893 & 2.7596 & 2.7951\\ \hline
$\Delta$ (\%) & 1.8 & 9.2 & 7.1 & 3.3 & 3.8 & 3.2\\ \hline
\end{tabular}
\vspace{10pt}
\caption{Numerical parameters and convergence results for the $2D$ system, grouped by the initial conditions; note that $\boldsymbol{\zeta}_{0}=(x(0),y(0),p_{x}(0),p_{y}(0))$.}
\label{tab:numerical_results_2D}
\end{table}

\clearpage
\bibliography{references}

\end{document}